\documentclass[amsmath,amsfonts,prx,twocolumn,longbibliography]{revtex4-1}
\usepackage{graphicx}
\usepackage{epsfig}
\usepackage{bbm}
\usepackage{bm}
\usepackage{dcolumn}
\usepackage{amsmath}
\usepackage{amssymb}
\usepackage{color}
\usepackage[varg]{txfonts}

\newcommand{\beg}{\begin{equation}}
\newcommand{\en}{\end{equation}}

\newcommand{\br}{\mathbf r}

\newcommand \bel  {\begin{align}}
\newcommand \enl  {\end{align}}

\newcommand{\eps}{\epsilon}

\newcommand{\dg}{^\dagger}

\begin{document}

\title{Collective modes in terahertz field response of superconductors with paramagnetic impurities}

\author{Yantao Li}
\affiliation{Department of Physics, Kent State University, Kent, OH 44242, USA}

\author{Maxim Dzero}
\affiliation{Department of Physics, Kent State University, Kent, OH 44242, USA}

\begin{abstract}
We consider a problem of nonlinear response to an external electromagnetic radiation of conventional disordered superconductors which contain a small amount of weak magnetic impurities. We focus on the diffusive limit and use Usadel equation to analyze the collective excitations  and obtain the dispersion relations for the collective modes.  We determine the resonant frequency and dispersion of both amplitude and phase (Carlson-Goldman) modes for moderate strength of magnetic scattering. We find that the Carlson-Goldman and superconducting plasmon modes can only be excited at some finite value of the threshold momentum which increases with an increase in spin-flip scattering rate while the amplitude mode is diffusive and becomes strongly suppressed with the increase in spin-flip scattering. The value of the threshold momentum is determined by the distance between the two consecutive spin-flip scattering events. Furthermore, we also find that the superconducting plasmon mode becomes gapless in the presence of the pair breaking processes. Possible ways towards experimental verification of our results are also  discussed. 
\end{abstract}

\date{\today}

\maketitle
\section{Introduction}
Advances in the state-of-the-art near- and far-field optical techniques \cite{THz1,THz2,Shimano2012,Shimano2013,Shimano2014,THz3,THz4,THz5,salvador2024principles} have renewed interest in the problems related to the study of the dynamical properties of superconductors. In particular, there have been significant research efforts directed towards elucidating various dynamical properties of conventional single band and spatially uniform \cite{Bergeret2018,Sun2020,Fan2022}, spatially nonuniform \cite{Hammer2016}, multiband \cite{Anishchanka2007,Khodas2014} as well as unconventional ($d$-wave) superconductors \cite{Ohashi1997,Dahm1998,Ohashi2000,Sharapov2002,Muller2021,Repplinger2023,Mootz2024,fiore2023manipulating,Sellati2023,Gabriele2021,Gabriele2022,Chu2023,Balatsky2000,Khodas2014,Hirschfeld1992,Lee2023,Bill2003,Benfatto2023,
Lee2023}. Many if not all of the works cited above have built on and further developed the ideas which have been put forward during the earlier attempts to investigate the dynamical properties of conventional superconductors by studying the spectrum of their collective excitations 
both in ballistic $l\gg\xi$ and diffusive $l\ll \xi$ regimes (here $l$ is the mean-free path and $\xi$ is the coherence length) \cite{VolkovKogan1973,Schmid1975,Artemenko1975,Carlson1973,Carlson1975,Kulik1981,Fertig1990,Fertig1991}. 

One important avenue for the renewed research efforts was opened thanks to the inquires about the specific role of potential disorder in influencing various response functions which are relevant for describing superconductors driven out-of-equilibrium. Indeed, it is well-known that if one neglects spatial inhomogeneities which can appear in the case of strong potential disorder, than scattering  does not affect superconductivity in equilibrium (Anderson theorem) \cite{AndersonTheorem,AG1961,Balatsky-RMP}. However, whether the Anderson theorem holds when a superconductor is driven out-of-equilibrium turns out to be a non-trivial question. 
Indeed, using quite broad variety of theoretical techniques (both numerical and analytical) several groups have recently given a special attention to the question of how disorder affects the transport and spectroscopic properties of out-of-equilibrium superconductors, especially in the context of the dynamics of amplitude Higgs mode \cite{Moore2017,Silaev2019,Samanta2020,Samanta2022,Yang2022-Disorder,Yang2020-Disorder2,Seibold2021,Cea2014-Disorder1,Derendorf2024}.   

\begin{figure}
\includegraphics[width=0.85\linewidth]{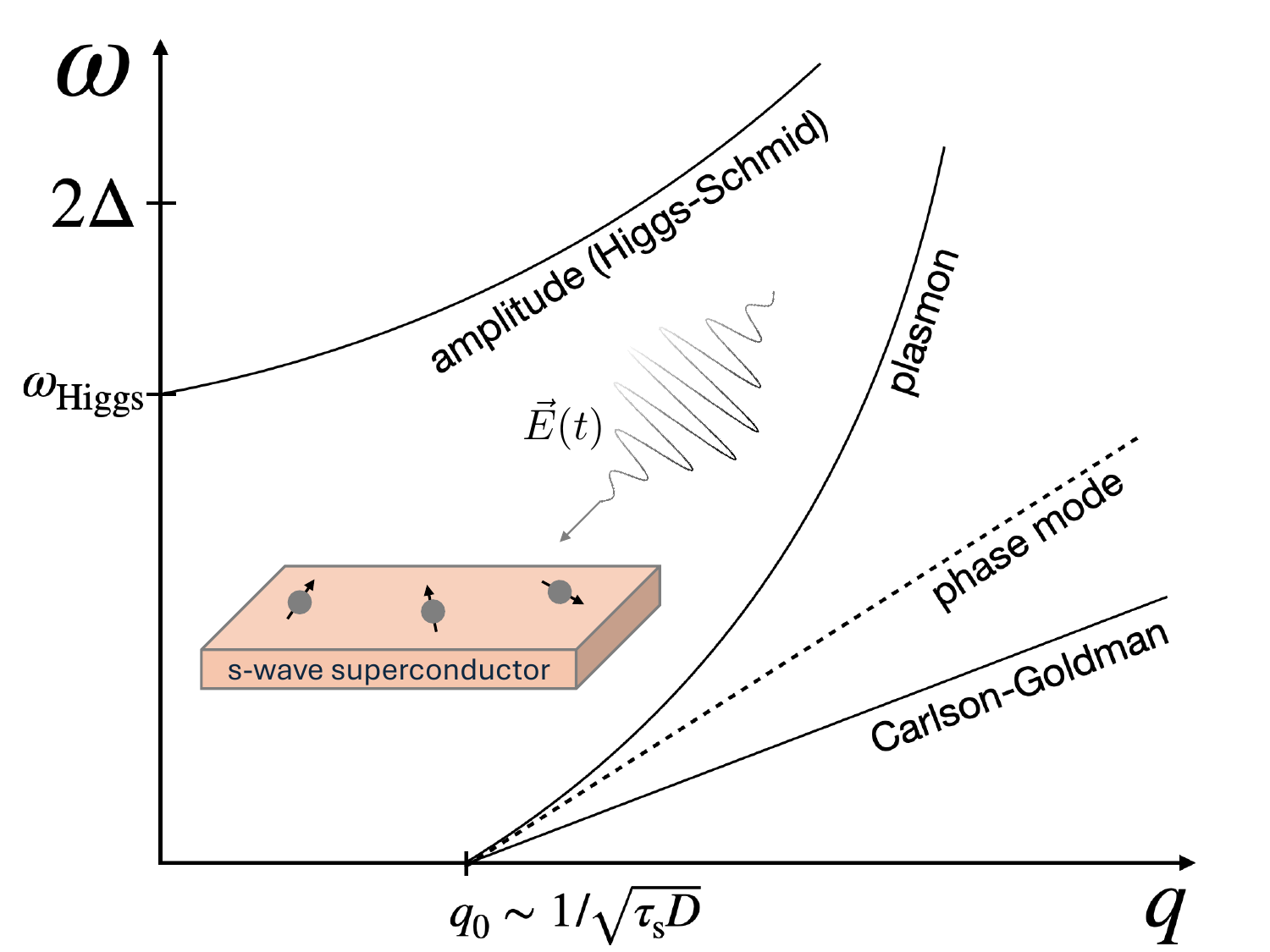}
\caption{Schematic diagram for the dispersion of the collective excitations produced by an external electromagnetic pulse in disordered conventional superconductor with paramagnetic impurities.  The momentum is assigned to the horizontal axis and frequency is assigned to the vertical axis. In the presence of the pair breaking scattering due to weak paramagnetic impurities the excitation of the superconducting plasmon and Calrson-Goldman (phase) modes can only take place at finite momentum. In a diffusive superconductor the value of the threshold momentum $q_0$ is determined by the pair breaking rate $1/\tau_{\textrm{s}}$ and the diffusion coefficient $D$. Pair breaking processes also lead to the red shift in the frequency of the amplitude Higgs-Schmid mode and render the superconducting plasmon mode gapless.}
\label{Fig-Main}
\end{figure}

It may seem surprising that despite its long history one generic aspect of this problem, namely the effects of the pair breaking disorder, such as paramagnetic impurities \cite{AG1961} or strong spatial inhomogeneities due to potential disorder \cite{LO-Inhomo}, on collective modes in conventional superconductors have not been discussed thus far within the framework of the microscopic theory. 
It is worth mentioning that in their pioneering work Gor'kov and Eliashberg have formulated a theory of nonstationary phenomena in disordered superconductors and applied it to several non-stationary problems, but did not discuss the problem of the collective modes \cite{LPG1968}. Most recently several studies have appeared which address the question of how scattering on weak paramagnetic impurities influences \emph{(i)} time evolution of the amplitude Higgs mode following the sudden perturbation \cite{Dzero2023-Disorder} and \emph{(ii)} frequency of the resonant amplitude mode, i.e. how it changes with an increase in spin-flip scattering rate \cite{Yantao2023}. However, to the best of our knowledge, the effects of weak paramagnetic impurities on other collective excitations in conventional superconductors have not been examined to date. 

Our goal in this paper is to try to fill this void. Specifically, we consider a diffusive superconductor contaminated with a small amount of weak paramagnetic impurities subjected to an external monochromatic electromagnetic field. Using the assumption that the scattering potential of paramagnetic impurities is weak we will ignore the effects associated with the existence of the Yu-Shiba-Rusinov bound states \cite{Marchetti2002,Fominov2011,Kharitonov2012}. We employ the Usadel equation which can be derived using the Keldysh field theoretical technique by performing an averaging over potential disorder \cite{Kamenev2009,KamenevBook}. Treating an external field as perturbation we compute the corrections to the superconducting order parameter and evaluate the dispersion of the collective modes associated with the order parameter's amplitude and phase fluctuations.  We also consider the effects associated with the Coulomb interactions by computing the correction to the Poisson equation which appears due to the charge re-distribution and obtain the dispersion of the superconducting plasmon mode. Our main results are schematically summarized in Fig. \ref{Fig-Main}.

\section{Qualitative discussion}
Under the action of the external terahertz (THz) electromagnetic field two collective modes, which can be attributed to the fluctuations of the phase and amplitude of the superconducting order parameter, will be excited in a superconductor \cite{Carlson1973,Artemenko1975,Schmid1975,Carlson1975,Kulik1981}. The excitation of the amplitude mode corresponds to the excitation of the Cooper pairs \cite{Varma2014} and in the conventional superconductor the minimum energy required to excite a Cooper pair is $\omega=2\Delta$, where $\Delta$ is the pairing gap. Conversely, the energy threshold to excite a single particle is $\omega_{\textrm{1p}}=\Delta$. In the presence of the pair breaking scattering due to paramagnetic impurities (or {spin-orbit} scattering) described by the relaxation time $\tau_{\textrm{s}}$, the threshold for the single particle excitation is decreased to a value $\omega_{\textrm{1p}}=\Delta[1-1/(\tau_{\textrm{s}}\Delta)^{2/3}]^{3/2}$ \cite{AG1961}. One therefore may expect that the pair breaking scattering should also lead to a softening of the threshold frequency for the excitation of the amplitude Higgs mode. Our earlier results \cite{Yantao2023} which have been based on the perturbative treatment of the spin-flip processes, confirm this expectation. Interestingly, this shift in the resonant frequency of the Higgs mode along with the smearing of the square-root singularity in the single-particle density of states suggests that the collisionless evolution of the order parameter becomes non-dissipative following a sudden perturbation induced by a short optical pulse \cite{Dzero2023-Disorder}.

In contrast with the amplitude mode, the phase mode is gapless in superconductors without pair breaking and, as a consequence, the energy of the phase mode is determined by its momentum $q$, $\omega\sim q$. In the presence of the pair breaking described by the scattering rate $\tau_{\textrm{s}}$, one may introduce a length scale $l_{\textrm{s}}=v_F\tau_{\textrm{s}}$, which represents a typical distance between the two consecutive spin-flip processes,  and associated with it momentum $q_{\textrm{s}}=\hbar/l_{\textrm{s}}$. As it turns out, the existence of $q_{\textrm{s}}$ implies that there should be a threshold value of momentum for the phase modes to be excited. In order to see this, it is convenient to use the Anderson pseudospins representation \cite{Anderson1959}. Anderson pseudospins operate in the phase space restricted to empty and doubly occupied states only and so the uniform rotation of all pseudospins at the corresponding single-particle energy levels $\eps$ around an axis defined by a vector ${\mathbf b}_\eps=(-\Delta,0,\eps)$ does not cost any energy, which means $\omega\to 0$ as $q\to 0$. Any spatially inhomogeneous rotation of the pseudospins in a disordered superconductor carries an energy cost $\omega\propto({\tau\Delta})^{1/2}v_Fq$ associated with it (here $\tau$ is the relaxation time due to scattering on potential impurities). It is clear that by construction pair breaking processes cause the destruction of the pseudospins. This implies that in the presence of weak paramagnetic impurities, the phase mode is only well defined on the length scales shorter than the pair breaking length scale, $l\leq l_{\textrm{s}}$. This, in turn, means that the phase mode can only be excited with momenta which satisfy $q\geq q_{\textrm{s}}$. Our subsequent macroscopic calculation presented below will indeed prove this conjecture.

Lastly, one should not forget that since the phase mode necessarily requires the breaking of the particle-hole symmetry, the excitation of the phase mode leads to the excitation of the plasmon mode. In a disordered superconductor without pair breaking this mode is gapped due to the finite energy cost for the excitation of this mode: $\omega_{\textrm{p}}\approx\sqrt{8\pi^2e^2\nu_FD\Delta_0}$. \cite{KamenevBook}. In the presence of the pair breaking not only do we find that this mode can only be excited with finite momentum, but also that this mode becomes gapless, Fig. \ref{Fig-Main}. We will limit our subsequent discussion to the case of very low temperatures $T\ll \Delta_0$.

\section{Main equations}
The main object of our subsequent analysis will be the $Q$-matrix 
\beg\label{QMatrix}
\check{Q}=\left(\begin{matrix} \hat{Q}^R & \hat{Q}^K \\ 0 & \hat{Q}^A\end{matrix}\right),
\en
where each component of $\check{Q}(\br;t,t')$ acts in Nambu and spin spaces \cite{Kamenev1999}. These matrices appear as a result of the Hubbard-Stratonovich transformation taken on the action which describes the BCS model \cite{BCS} after the averaging over all potential disorder configurations has been performed under an assumption that the disorder distribution is Gaussian \cite{Kamenev1999,Kamenev2009,KamenevBook}. 
The saddle point $Q$-matrix configuration of the effective action is given by the solution of the Usadel equation which describes a conventional strongly disordered superconductor with weak paramagnetic impurities:
\beg\label{UsadelEq}
\begin{split}
&\check{\Xi}_3\partial_t\check{Q}+\partial_{t'}\check{Q}\check{\Xi}_3-\hat{\partial}_\br\left(D\check{Q}\circ\hat{\partial}_\br\check{Q}\right)-i[\check{\Delta},\check{Q}]+i[\check{\Phi},\check{Q}]\\&+\frac{1}{6\tau_{\textrm{s}}}[\check{\Gamma}_k\check{Q}\check{\Gamma}_k\stackrel{\circ}{,}\check{Q}]=0.
\end{split}
\en
Here $\hat{\gamma}^{\textrm{cl}}$ is a unit matrix acting in Keldysh space, $\check{\Xi}_3=\hat{\gamma}^{\textrm{cl}}\otimes\hat{\Xi}_3$ is diagonal in Keldysh space, $\check{\Gamma}_k=\hat{\gamma}^{\textrm{cl}}\otimes\hat{\Gamma}_k$ and $\check{\Phi}=\Phi(\br,t)\check{\Xi}_0$ is a matrix scalar potential which should also be understood as diagonal in Nambu and Keldysh spaces. We have introduced the following matrices $\hat{\Gamma}_k=\hat{\rho}_3\otimes\hat{\sigma}_k$, $\hat{\Xi}_2=i\hat{\rho}_2\otimes\hat{\sigma}_0$, $\hat{\Xi}_1=\hat{\rho}_1\otimes\hat{\sigma}_0$ and $\hat{\Xi}_3=\hat{\rho}_3\otimes\hat{\sigma}_0$,
where $\hat{\rho}_a$ and $\hat{\sigma}_b$ are Pauli matrices. 
Superconducting order parameter matrix $\check{\Delta}(\br,t)$ is defined 
according to
\beg\label{DLTMatrix}
\begin{split}
\check{\Delta}(\br,t)&=\Delta(\br,t)\left(\hat{\rho}_{+}\otimes\hat{\sigma}_0\right)-\overline{\Delta}(\br,t)\left(\hat{\rho}_{-}\otimes\hat{\sigma}_0\right),
\end{split}
\en
where $\hat{\rho}_{\pm}=\hat{\rho}_1\pm i\hat{\rho}_2$ and $\check{\Delta}$ is diagonal in Keldysh space. Order parameter is determined self-consistently from 
\beg\label{SelfConsistent}
\Delta(\br,t)=\frac{\pi\lambda}{2}\textrm{Tr}\left\{\left(\hat{\gamma}^{\textrm{q}}\otimes\hat{\rho}_{-}\otimes\hat{\sigma}_0\right)
\otimes\check{Q}(\br;t,t)\right\},
\en
where $\hat{\gamma}^{\textrm{q}}$ is the first Pauli matrix acting in the Keldysh subspace, $\lambda$ is the dimensionless coupling constant. 
Lastly, the covariant derivative is defined according to
\beg\label{CoDeriv}
\hat{\partial}_\br\check{Q}=\partial_\br\check{Q}-i\left[\check{\Xi}_3{\mathbf A},\check{Q}\right]=
\partial_\br\check{Q}-i{\mathbf A}(t)\check{\Xi}_3\check{Q}+i\check{Q}{\mathbf A}(t')\check{\Xi}_3
\en 
and ${\mathbf A}={\vec A}(\br,t)$ is a vector potential.  

\begin{figure}
\includegraphics[width=0.85\linewidth]{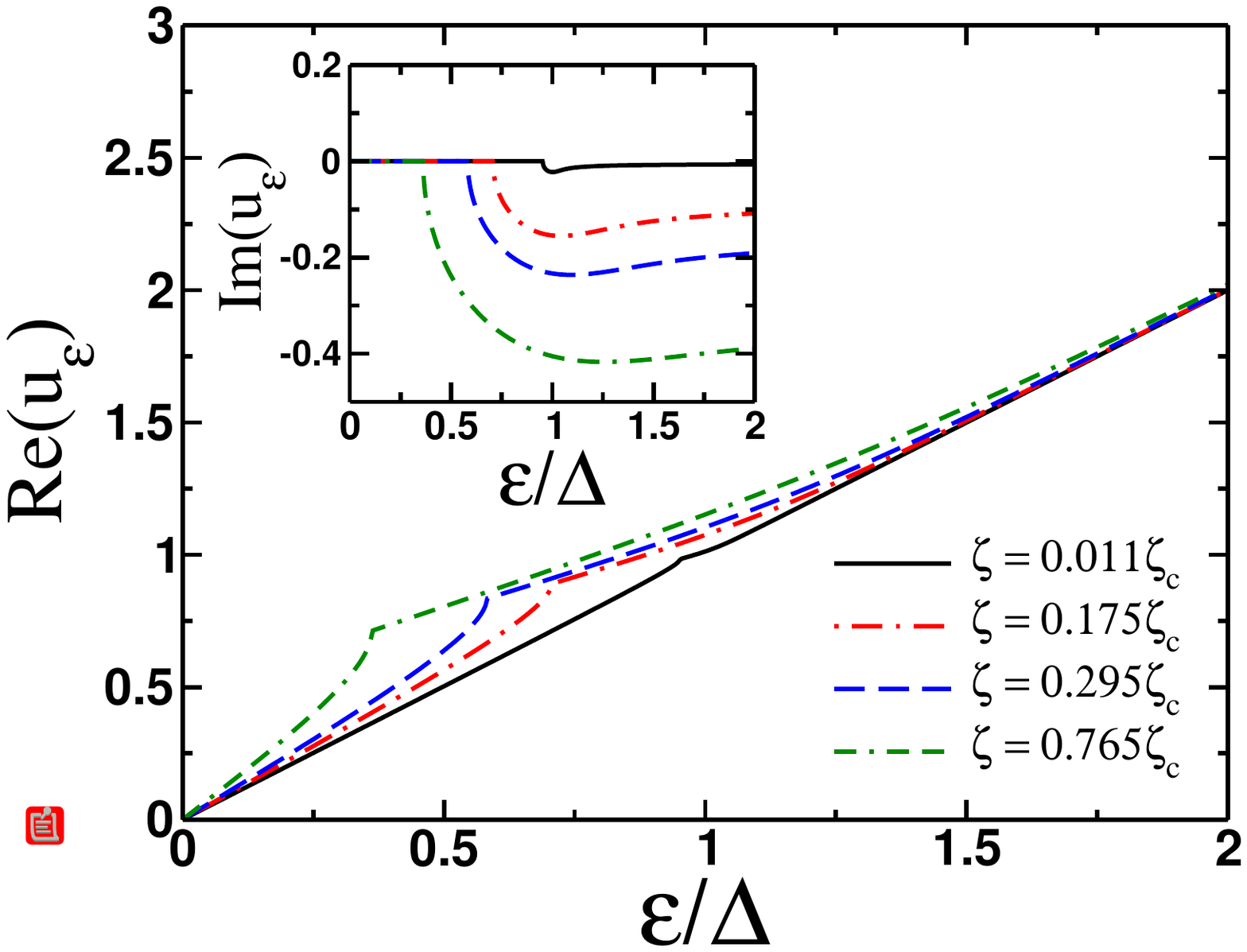}
\caption{Results of the numerical solution of the equation (\ref{Eq4ueps}) for real and imaginary parts of $u_\eps$ as a function of $\eps/\Delta$ for various values of the dimensionless parameter $\zeta=1/\tau_{\textrm{s}}\Delta_0$. Here $\Delta_0=\Delta(\zeta=0)$ is the pairing gap in the absence of paramagnetic disorder, $\zeta_c$ denotes the critical value of $\zeta$ when the order parameter vanishes, $\Delta(\zeta_c)=0$.}
\label{Fig-ueps}
\end{figure}

\subsection{Ground state} 
In the ground state $Q$-matrix is spatially homogeneous, $\check{Q}_{\eps\eps'}=2\pi\check{\Lambda}_\eps\delta(\eps-\eps')$, and the Keldysh component is just parametrization $\hat{\Lambda}_\eps^K=(\hat{\Lambda}_\eps^R-\hat{\Lambda}_\eps^A)\tanh(\eps/2T)$ to ensure the normalization condition $\check{\Lambda}_\eps^2=\check{\mathbbm{1}}$ (here $T$ is temperature).
By looking at the Usadel equation it becomes clear that we can search for the solution for $\hat{Q}^{R(A)}$ in the following form:
\beg\label{AnsatzGR}
\hat{\Lambda}_\eps^{R}=(\hat{\rho}_3\otimes\hat{\sigma}_0)g_\eps^{R}+(i\hat{\rho}_2\otimes\hat{\sigma}_0)f_\eps^{R}
\en
and we choose $\hat{\Delta}=\left(i\hat{\rho}_2\otimes\hat{\sigma}_0\right)\Delta$.
The advanced component of $\check{\Lambda}_{\eps}$ can be found from
$\hat{\Lambda}_\eps^A=-\hat{\Xi}_3\left(\hat{\Lambda}_\eps^R\right)\dg\hat{\Xi}_3$.
After we insert (\ref{AnsatzGR}) into the Usadel equation (\ref{UsadelEq}), for the components (\ref{AnsatzGR}) we find
$g_\eps^R=u_\eps\textrm{sign}(\eps)/\sqrt{u_\eps^2-1}$ and $f_\eps^R=\textrm{sign}(\eps)/\sqrt{u_\eps^2-1}$, where $u_\eps$ is given by the solution of the following nonlinear equation
\beg\label{Eq4ueps}
u_\eps\left(1+\frac{1}{\tau_{\textrm{s}}\Delta}\frac{\sqrt{1-u_\eps^2}}{u_\eps^2-1}\right)=\frac{\eps}{\Delta},
\en
and $\Delta$ is computed self-consistently using Eq. (\ref{SelfConsistent}) and the expressions above.
In Fig. \ref{Fig-ueps} we show the dependence of $u_\eps$ on $\eps/\Delta$ computed for various values of $1/\tau_{\textrm{s}}\Delta$.
\subsection{Linearized Usadel equation}
All our subsequent analysis will be focused on computing the linear correction to the equilibrium $Q$-matrix configuration, so we write
\beg\label{LinCorr}
\check{Q}_{\eps\eps'}=2\pi\check{\Lambda}_\eps\delta(\eps-\eps')+\delta\check{G}_{\eps\eps'}.
\en
In what follows we assume that the external vector potential is spatially homogeneous and is given by the superposition of two monochromatic waves with frequencies $\Omega_\nu$ and $\Omega_\mu$:
\beg\label{Anm}
{\mathbf A}(t)={\mathbf A}_\nu e^{i\Omega_\nu t}+{\mathbf A}_\mu e^{i\Omega_\mu t}+
{\mathbf A}_\nu^* e^{-i\Omega_\nu t}+{\mathbf A}_\mu^* e^{-i\Omega_\mu t}.
\en
Treating an external vector potential as perturbation, we find the following linearized equation for the matrix function $\delta\check{G}_{\eps\eps'}(q)$:
\beg\label{LinUsadel}
\begin{split}
&\left(\eps\check{\Xi}_3+\check{\Delta}+\frac{i}{6\tau_{\textrm{s}}}\sum\limits_{a=1}^3\check{\Gamma}_a\check{\Lambda}_\eps\check{\Gamma}_a\right)\delta\check{G}_{\eps\eps'}\\&-\delta\check{G}_{\eps\eps'}
\left(\eps'\check{\Xi}_3+\check{\Delta}+\frac{i}{6\tau_{\textrm{s}}}\sum\limits_{a=1}^3\check{\Gamma}_a\check{\Lambda}_\eps\check{\Gamma}_a\right)-iD\check{\Lambda}_\eps{\partial}_\br^2\delta\check{G}_{\eps\eps'}\\&=2\pi\sum\limits_{\mu\nu}\left[\check{R}_{\cal A}(\eps,\eps')+\check{R}_{\Delta}(\eps,\eps')\right]\delta(\eps-\eps'-\Omega_{\nu+\mu})\\&+\frac{i}{6\tau_{\textrm{s}}}
\sum\limits_{a=1}^3\left(\check{\Lambda}_\eps\check{\Gamma}_a\delta\check{G}_{\eps\eps'}\check{\Gamma}_a-\check{\Gamma}_a\delta\check{G}_{\eps\eps'}\check{\Gamma}_a\check{\Lambda}_{\eps'}\right)\\&+\check{\Phi}_{\eps-\eps'}\check{\Lambda}_{\eps'}-\check{\Lambda}_\eps\check{\Phi}_{\eps-\eps'}.
\end{split}
\en
Here $\Omega_{\nu+\mu}$ is given by four possible combinations of $\pm\Omega_{\nu}$ and $\pm\Omega_\mu$, i.e. $\Omega_{\nu+\mu}=\pm(\Omega_\nu\pm\Omega_\mu)$. Here function $\check{R}_{\Delta}(\eps,\eps')$ represents the correction to the order parameter
\beg\label{RDelta}
\check{R}_{\Delta}(\eps,\eps')=\check{\Lambda}_\eps\delta\check{\Delta}(\eps-\eps')-\delta\check{\Delta}(\eps-\eps')\check{\Lambda}_{\eps'}.
\en
Function $\check{R}_{\cal A}(\eps,\eps')$ is of the second order in powers of the vector potential
\beg\label{RA}
\check{R}_{\cal A}(\eps,\eps')=iD{\cal T}_{\mu\nu}\left(\check{\Lambda}_\eps\check{\Xi}_3\check{\Lambda}_{\eps'-\Omega_\mu}\check{\Xi}_3-\check{\Xi}_3\check{\Lambda}_{\eps+\Omega_\nu}\check{\Xi}_3\check{\Lambda}_{\eps'}\right),
\en
where ${\cal T}_{\mu\nu}={\mathbf A}_\nu{\mathbf A}_\mu$ determines the intensity of external electromagnetic field. The details of the subsequent analysis of this equation depend on the specific choice of the matrix structure for $\delta\check{G}_{\eps\eps'}(q)$ since it will determine the nature of the excitations, i.e. phase or amplitude modes.  
\section{Anderson-Bogoliubov mode}
We start with the analysis of the phase mode in a charge neutral case $(e=0)$. Since the excitation of the phase mode requires a breaking of the particle-hole symmetry \cite{Sun2020} and the phase mode fluctuations are connected with the particle number fluctuations, the normal component of $\delta\check{G}$ must be proportional to the unit matrix, while the pairing component as well as linear correction to the order parameter must be proportional to the matrix $\hat{\Xi}_1$. Hence we write
\beg\label{dGAB}
\delta\hat{G}_{\eps\eps'}^{(\alpha)}(q)=\delta G_{\eps\eps'}^{(\alpha)}(q)\hat{\Xi}_0+\delta F_{\eps\eps'}^{(\alpha)}(q)\hat{\Xi}_1, \quad \delta\hat{\Delta}=i\delta\Delta_\omega^T(q)\hat{\Xi}_1,
\en
where the superscript $\alpha$ refers to the retarded, advanced and Keldysh components and $\omega=\eps-\eps'$. 
\paragraph{Perturbative solution.} Let us limit ourselves to the discussion of the approximate solution by neglecting the terms proportional to $1/\tau_{\textrm{s}}$ in the right hand side of the {linearized} Usadel equation (\ref{LinUsadel}) i.e. we will consider the case when $1/\tau_{\textrm{s}}\Delta\ll 1$. Then for the retarded and advanced components of $\delta\check{G}$ we have
\beg\label{ApproxSolAB}
\begin{split}
&\delta\hat{G}_{\eps\eps'}^R(q)\approx\frac{\delta\hat{\Delta}-\hat{\Lambda}_{\eps}^R\delta\hat{\Delta}\hat{\Lambda}_{\eps'}^R}{\eta_\eps^R+\eta_{\eps'}^R}\left(1-\frac{iDq^2}{\eta_\eps^R+\eta_{\eps'}^R}\right), \\ 
&\delta\hat{G}_{\eps\eps'}^A(q)\approx\frac{\delta\hat{\Delta}-\hat{\Lambda}_{\eps}^A\delta\hat{\Delta}\hat{\Lambda}_{\eps'}^A}{\eta_\eps^A+\eta_{\eps'}^A}\left(1-\frac{iDq^2}{\eta_\eps^A+\eta_{\eps'}^A}\right).
\end{split}
\en
Here we explicitly assumed that momentum is small, $Dq^2\ll \Delta$, introduced notations $\eta_\eps^{R(A)}=\textrm{sign}(\eps)\tilde{\Delta}_{\eps}^{R(A)}\sqrt{u_\eps^2-1}$
and 
\beg\label{tDLTRA}
\tilde{\Delta}_{\eps}^{R(A)}=\Delta-\left(\frac{i}{2\tau_{\textrm{s}}}\right)f_\eps^{R(A)}.
\en
It is customary to look for the Keldysh component as a sum of the regular and anomalous contributions:
\beg\label{dGKAB}
\delta\hat{G}_{\eps\eps'}^K(q)=\delta\hat{G}_{\textrm{reg}}^K(q;\eps,\eps')+\delta\hat{G}_{\textrm{an}}^K(q;\eps,\eps').
\en
The regular contribution is given in terms of the corrections to the retarded and advanced components (\ref{ApproxSolAB}):
\beg\label{gKKABreg}
\delta\hat{G}_{\textrm{reg}}^K(q;\eps,\eps')=\delta\hat{G}_{\eps\eps'}^R(q)t_{\eps'}-t_\eps\delta\hat{G}_{\eps\eps'}^A(q),
\en
where $t_\eps=\tanh(\eps/2T)$. After a somewhat tedious calculation, we found the following expressions for the components $\delta\hat{G}_{\textrm{an}}^K(q;\eps,\eps')$:
\beg\label{dGFanFin}
\begin{split}
\delta G_{\textrm{an}}^K(\eps,\eps';q)&=\left\{\frac{(t_\eps-t_{\eps'}){\cal B}_{\eps\eps'}^K}{\eta_\eps^R+\eta_{\eps'}^A}\left(1-\frac{iDq^2}{\eta_\eps^R+\eta_{\eps'}^A}\right)\right.\\&\left.-\frac{iDq^2t_\eps\left({\cal A}_{\eps\eps}^K{\cal B}_{\eps\eps'}^A-{\cal B}_{\eps\eps}^K{\cal A}_{\eps\eps'}^A\right)}{(\eta_\eps^R+\eta_{\eps'}^A)(\eta_\eps^A+\eta_{\eps'}^A)}\right\}i\delta\Delta_\omega^T, \\
\delta F_{\textrm{an}}^K(\eps,\eps';q)&=\left\{\frac{(t_\eps-t_{\eps'}){\cal A}_{\eps\eps'}^K}{\eta_\eps^R+\eta_{\eps'}^A}\left(1-\frac{iDq^2}{\eta_\eps^R+\eta_{\eps'}^A}\right)\right.\\&\left.-\frac{iDq^2t_\eps\left({\cal A}_{\eps\eps}^K{\cal A}_{\eps\eps'}^A-{\cal B}_{\eps\eps}^K{\cal B}_{\eps\eps'}^A\right)}{(\eta_\eps^R+\eta_{\eps'}^A)(\eta_\eps^A+\eta_{\eps'}^A)}\right\}i\delta\Delta_\omega^T.
\end{split}
\en
The expressions for the functions ${\cal A}_{\eps\eps'}^{R,A,K}$ and ${\cal B}_{\eps\eps'}^{R,A,K}$ are listed in the Appendix \ref{AppendixA}.
\begin{figure}
\includegraphics[width=0.85\linewidth]{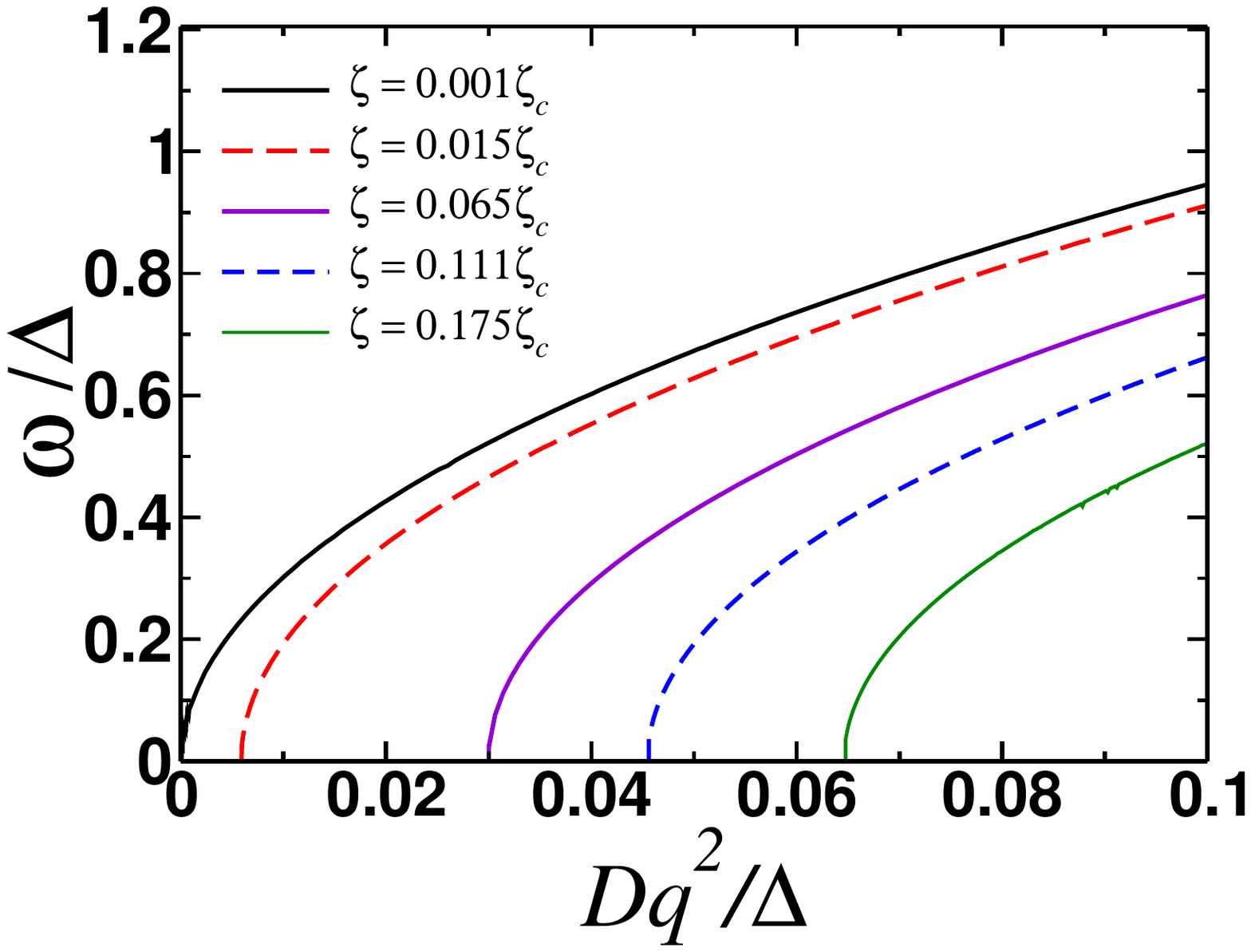}
\caption{Dispersion of the Anderson-Bogoliubov mode at very low temperatures $T\approx 10^{-3}\Delta_0$, where $\Delta_0$ is the pairing gap in a clean superconductor. We also find that $\omega\propto q$. Here $\Delta$ is a superconducting order parameter computed for finite values of disorder $\zeta$. The critical value of $\zeta$ is determined by $2/(\tau_{\textrm{s}}^{(c)}\Delta_0)=1$ or $\zeta_{\textrm{c}}=0.855$ for our choice for the values of the coupling constant and ultraviolet cutoff.}
\label{Fig-AB}
\end{figure}

\paragraph{Dispersion relation.}
In order to determine the dispersion (and damping) of the transverse mode, we use the {linearized} self-consistency condition
\beg\label{Self1}
i\delta\Delta_\omega^T(q)=\frac{\lambda}{2}\int\limits_{-\infty}^\infty \delta F_{\eps,\eps-\omega}^K(q)d\eps.
\en
In what follows we will also need to use the self-consistency condition for the ground state
\beg\label{Invlam}
\frac{2}{\lambda}=\int\limits_{-\omega_D}^{\omega_D}d\eps\left(\frac{f_\eps^R}{\Delta}-\frac{f_\eps^A}{\Delta}\right)t_\eps.
\en
Using the expressions above we recast (\ref{Self1}) into the following form:
\beg\label{Recast}
\left[a(\omega)-Dq^2b(\omega)\right]i\delta\Delta_\omega^T(q)=0,
\en
where we introduced functions
\beg\label{abw}
\begin{split}
a(\omega)&=\frac{1}{2}\int\limits_{-\infty}^\infty d\eps \left[\frac{(t_\eps-t_{\eps'}){\cal A}_{\eps\eps'}^K}{\eta_\eps^R+\eta_{\eps'}^A}+
\frac{{\cal A}_{\eps\eps'}^Rt_{\eps'}}{\eta_\eps^R+\eta_{\eps'}^R}
\right.\\&\left.-
\frac{t_\eps{\cal A}_{\eps\eps'}^A}{\eta_\eps^A+\eta_{\eps'}^A}-\left(\frac{f_\eps^R}{\Delta}-\frac{f_\eps^A}{\Delta}\right)t_\eps\right], \\
b(\omega)&=\frac{i}{2}\int\limits_{-\infty}^\infty d\eps \left[\frac{(t_\eps-t_{\eps'}){\cal A}_{\eps\eps'}^K}{(\eta_\eps^R+\eta_{\eps'}^A)^2}+
\frac{{\cal A}_{\eps\eps'}^Rt_{\eps'}}{(\eta_\eps^R+\eta_{\eps'}^R)^2}\right.\\&\left.-
\frac{t_\eps{\cal A}_{\eps\eps'}^A}{(\eta_\eps^A+\eta_{\eps'}^A)^2}+\frac{t_\eps\left({\cal A}_{\eps\eps}^K{\cal A}_{\eps\eps'}^A-{\cal B}_{\eps\eps}^K{\cal B}_{\eps\eps'}^A\right)}{(\eta_\eps^R+\eta_{\eps'}^A)(\eta_\eps^A+\eta_{\eps'}^A)}\right]
\end{split}
\en
and $\eps'=\eps-\omega$. As we have already mentioned in the introductory part of this paper, $\omega=q=0$ is not a solution of equation (\ref{Recast}). Indeed, setting $\omega=q=0$ in (\ref{Recast}) and using ${\cal A}_{\eps\eps}^{R(A)}=2$ (Appendix A) we can cast equation (\ref{Recast}) into the following simple form
\beg\label{wq0}
\begin{split}
&\int\limits_{-\infty}^\infty d\eps\left[\frac{1}{\eta_\eps^R}\left(1-\frac{\tilde{\Delta}_\eps^R}{\Delta}\right)
-\frac{1}{\eta_\eps^A}\left(1-\frac{\tilde{\Delta}_\eps^A}{\Delta}\right)\right]t_{\eps}\\&=\frac{i}{2\tau_{\textrm{s}}\Delta}\int\limits_{-\infty}^\infty d\eps\left(\frac{f_\eps^R}{\eta_\eps^R}-\frac{f_\eps^A}{\eta_\eps^A}\right)t_\eps.
\end{split}
\en
Clearly, this expression is zero only in the clean limit $\tau_{\textrm{s}}\to \infty$. Therefore, we discover that the phase mode can only be excited for the finite values of momenta $q$. The threshold value $q_{\textrm{s}}$ is found from
\beg\label{qth}
b(0)Dq_{\textrm{s}}^2=\frac{i}{2\tau_{\textrm{s}}\Delta}\int\limits_{-\infty}^\infty d\eps\left(\frac{f_\eps^R}{\eta_\eps^R}-\frac{f_\eps^A}{\eta_\eps^A}\right)t_\eps.
\en 
For $\tau_{\textrm{s}}\Delta\gg 1$ the threshold momentum $q_{\textrm{s}}\sim\sqrt{1/D\tau_{\textrm{s}}}\sim\hbar/v_F\tau_{\textrm{s}}$. Therefore $q_{\textrm{s}}$ is determined by the distance $l_s=v_F\tau_{\textrm{s}}$ between two consecutive spin-flip scattering events. In Fig. \ref{Fig-AB} we show the result of the numerical solution of the equation (\ref{Recast}) for the dispersion of the phase mode $\omega(q)$. 
The dispersion of the transverse mode is computed numerically and the results are presented in Fig. \ref{Fig-AB}. The dispersion relation can be approximately described by 
\beg\label{AB-wq}
\omega_{\textrm{AB}}(q)\propto v_F(q-q_{\textrm{s}})\vartheta(q-q_{\textrm{s}}),
\en
where $\vartheta(x)$ is the Heaviside step-function. For the results shown in Fig. \ref{Fig-AB} the damping of the transverse mode has been neglected. However, when we have included the damping, from numerical calculations for the damping rate $\omega-i\gamma$ we find that $\gamma\propto Dq^2/\Delta$ and it also increases with $1/\tau_{\textrm{s}}\Delta_0$. 

\section{Carlson-Goldman mode}
Let us {now} relax the assumption from the previous section and consider the case of charged superfluid (superconductor), $e\not =0$. We will still use the ansatz (\ref{dGAB}),
but the self-consistency condition for the order parameter must now be supplemented with the Poisson equation:
\beg\label{Poisson}
-\frac{q^2}{4\pi e^2}\Phi_\omega(q)=2\nu_F\Phi_\omega(q)+\frac{\nu_F}{2}\int\limits_{-\infty}^{+\infty}{d\eps}\delta G_{\eps,\eps-\omega}^K(q).
\en
Here $\nu_F$ is the single particle density of states at the Fermi level and the first term in the right hand side of this equation appears due to static polarizability of the conduction band \cite{KamenevBook}.
We need to emphasize here that given our choice of the ground state configuration (\ref{AnsatzGR}), the terms which contain the vector potential will not contribute to $\delta\hat{G}_{\eps\eps'}(q)$. On the other hand, if we were to include in (\ref{AnsatzGR}) the term which is proportional to $\hat{\Xi}_1$, which is equivalent to having a complex order parameter in the ground state, than electromagnetic field will directly contribute to both $\Phi_\omega(q)$ and $\delta\Delta_\omega^T(q)$ leading to the excitation of the phase mode. 
For the sake of simplicity, we will continue working assuming that the order parameter is real in the ground state with the understanding that $\delta\Delta_\omega^T(q)$ can be induced by an external terahertz radiation \cite{Sun2020}.

After a quick calculation, we find the following expressions for the correction to the retarded and advanced components which determine the regular correction to the Keldysh function $\delta\hat{G}_{\textrm{reg}}^K$, Eq. (\ref{gKKABreg}):
\beg\label{CompsReg}
\begin{split}
&\delta{G}_{\eps\eps'}^{R(A)}(q)=\frac{{\cal B}_{\eps\eps'}^{R(A)}i\delta\Delta_\omega^T-\overline{\cal A}_{\eps\eps'}^{R(A)}\Phi_\omega}{\eta_\eps^{R(A)}+\eta_{\eps'}^{R(A)}+iDq^2}, \\
&\delta{F}_{\eps\eps'}^{R(A)}(q)=\frac{{\cal A}_{\eps\eps'}^{R(A)}i\delta\Delta_\omega^T+{\cal B}_{\eps\eps'}^{R(A)}\Phi_\omega}{\eta_\eps^{R(A)}+\eta_{\eps'}^{R(A)}+iDq^2}.
\end{split}
\en
Here we retain the momentum terms $Dq^2$ in the denominators of these expressions for compactness. In what follows below we will expand these expressions in powers of $Dq^2$ up to the first order. 

Computation of $\delta\hat{G}_{\textrm{an}}^K$ can be done similarly to the one from the previous Section and it yields the following results:
\begin{widetext}
\beg\label{dKGanCC}
\begin{split}
&\delta G_{\textrm{an}}=\left\{\frac{(t_\eps-t_{\eps'}){\cal B}_{\eps\eps'}^K}{\eta_\eps^R+\eta_{\eps'}^A+iDq^2}-iDq^2\frac{t_\eps({\cal A}_{\eps\eps}^K{\cal  B}_{\eps\eps'}^A-{\cal B}_{\eps\eps}^K{\cal A}_{\eps\eps'}^A)}{(\eta_\eps^R+\eta_{\eps'}^A)(\eta_\eps^A+\eta_{\eps'}^A)}\right\}i\delta\Delta_\omega^T-\left\{\frac{(t_\eps-t_{\eps'})\overline{\cal A}_{\eps\eps'}^K}{\eta_\eps^R+\eta_{\eps'}^A+iDq^2}-iDq^2\frac{t_\eps({\cal A}_{\eps\eps}^K\overline{\cal  A}_{\eps\eps'}^A+{\cal B}_{\eps\eps}^K{\cal  B}_{\eps\eps'}^A)}{(\eta_\eps^R+\eta_{\eps'}^A)(\eta_\eps^A+\eta_{\eps'}^A)}\right\}\Phi_\omega, \\
&\delta F_{\textrm{an}}=\left\{\frac{(t_\eps-t_{\eps'}){\cal A}_{\eps\eps'}^K}{\eta_\eps^R+\eta_{\eps'}^A+iDq^2}-iDq^2\frac{t_\eps({\cal A}_{\eps\eps}^K{\cal  A}_{\eps\eps'}^A-{\cal B}_{\eps\eps}^K{\cal  B}_{\eps\eps'}^A)}{(\eta_\eps^R+\eta_{\eps'}^A)(\eta_\eps^A+\eta_{\eps'}^A)}\right\}i\delta\Delta_\omega^T
+\left\{\frac{(t_\eps-t_{\eps'}){\cal B}_{\eps\eps'}^K}{\eta_\eps^R+\eta_{\eps'}^A+iDq^2}-iDq^2\frac{t_\eps({\cal A}_{\eps\eps}^K{\cal  B}_{\eps\eps'}^A+{\cal B}_{\eps\eps}^K\overline{\cal  A}_{\eps\eps'}^A)}{(\eta_\eps^R+\eta_{\eps'}^A)(\eta_\eps^A+\eta_{\eps'}^A)}\right\}\Phi_\omega.
\end{split}
\en
\end{widetext}
Here again, we have retained $Dq^2$ in the denominators of these expressions for brevity. In order to make contact with the results of the earlier studies \cite{Kulik1981} let us consider the situation when the paramagnetic impurities are absent.  In the limit $\tau_{\textrm{s}}\to\infty$ expressions (\ref{CompsReg},\ref{dKGanCC}) can be significantly simplified (see Appendix B for details). In particular, in this case as it follows directly from equation (\ref{Eq4ueps}) for the single particle propagators we find $g_\eps^{R(A)}=\eps/\eta_\eps^{R(A)}$, $f_\eps^{R(A)}=\Delta_0/\eta_{\eps}^{R(A)}$ with
\beg\label{etaRA}
\eta_\eps^{R(A)}=\left\{
\begin{matrix} 
\pm\textrm{sign}(\eps)\sqrt{(\eps\pm i0)^2-\Delta_0^2}, & |\eps|\geq\Delta_0, \\
i\sqrt{\Delta_0^2-\eps^2}, & |\eps|<\Delta_0.
\end{matrix}\right.
\en
As a result, we obtain the following expressions for the corrections to the Keldysh function:
\beg\label{dGKPoten}
\begin{split}
&\left[\delta G_{\eps\eps'}^K\right]_{\tau_{\textrm{s}}\to\infty}=-\frac{\Phi_\omega}{\omega}\left[t_\eps(g_\eps^R-g_\eps^A)-t_{\eps'}(g_{\eps'}^R-g_{\eps'}^A)\right]\\&+\frac{(i\omega\delta\Delta_\omega^T-2\Delta \Phi_\omega)}{\omega(\eps+\eps')}[t_{\eps'}(f_{\eps'}^R-f_{\eps'}^A)-t_\eps(f_\eps^R-f_\eps^A)], \\
\end{split}
\en
and
\beg\label{dFKPoten}
\begin{split}
&\left[\delta F_{\eps\eps'}^K\right]_{\tau_{\textrm{s}}\to\infty}=\frac{t_\eps(g_\eps^R-g_\eps^A)+t_{\eps'}(g_{\eps'}^R-g_{\eps'}^A)}{\eps+\eps'}i\delta\Delta_\omega^T\\&-
\frac{t_\eps(f_\eps^R-f_\eps^A)-t_{\eps'}(f_{\eps'}^R-f_{\eps'}^A)}{\eps+\eps'}\Phi_\omega.
\end{split}
\en
Here we have ignored the terms $\propto Dq^2$ for brevity.
These expressions coincide with the corresponding expressions listed in Refs. \cite{Kulik1981,KamenevBook}. There is one important observation which can be made here: as it can be directly checked the integral over $\eps$ of the first term in the expression for
$\left[\delta G_{\eps\eps'}^K\right]_{\tau_{\textrm{s}}\to\infty}$ is actually independent of $\omega$ and it equals to $-4\Phi_\omega$ which means that the contribution from this term will exactly cancel out the corresponding conduction band polarizability term in the Poisson equation (\ref{Poisson}).

We now go back to the general case of finite $\tau_{\textrm{s}}$. After collecting all the contributions (\ref{CompsReg},\ref{dKGanCC}) together into expression (\ref{dGKAB}), the self-consistency equation (\ref{Self1}) can be cast into the following form
\beg\label{SelfCC}
[a(\omega)-Dq^2b(\omega)]\delta\Delta_\omega^T-[i\tilde{a}(\omega)+Dq^2\tilde{d}(\omega)]\Phi_\omega=0,
\en
where functions $\tilde{a}(\omega)$ and $\tilde{b}(\omega)$ are defined in Appendix A. Notably, in the limit $\omega\to 0$ both of these functions are zero as we immediately recover the corresponding equation for the minimum momentum threshold from the previous Section, Eq. (\ref{AB-wq}). Consequently, the Poisson equation (\ref{Poisson}) can be re-written as 
\beg\label{Re-Poisson}
[i\tilde{a}(\omega)+Dq^2\tilde{b}(\omega)]\delta\Delta_\omega^T+\left[s(\omega)-Dq^2{d}(\omega)\right]\Phi_\omega=0.
\en
It is straightforward to check that in the limit $\omega\to 0$ equations (\ref{SelfCC}) and (\ref{Re-Poisson}) remain coupled to each other since $\tilde{b}(0)\not=0$ and $s(0)\not=0$. By requiring that the determinant of the resulting system of linear equations vanishes we then recover our result for the momentum threshold, while equation (\ref{Re-Poisson}) provides a linear relation between $i\delta\Delta_\omega^T$ and $\Phi_\omega$.

\begin{figure}
\includegraphics[width=0.85\linewidth]{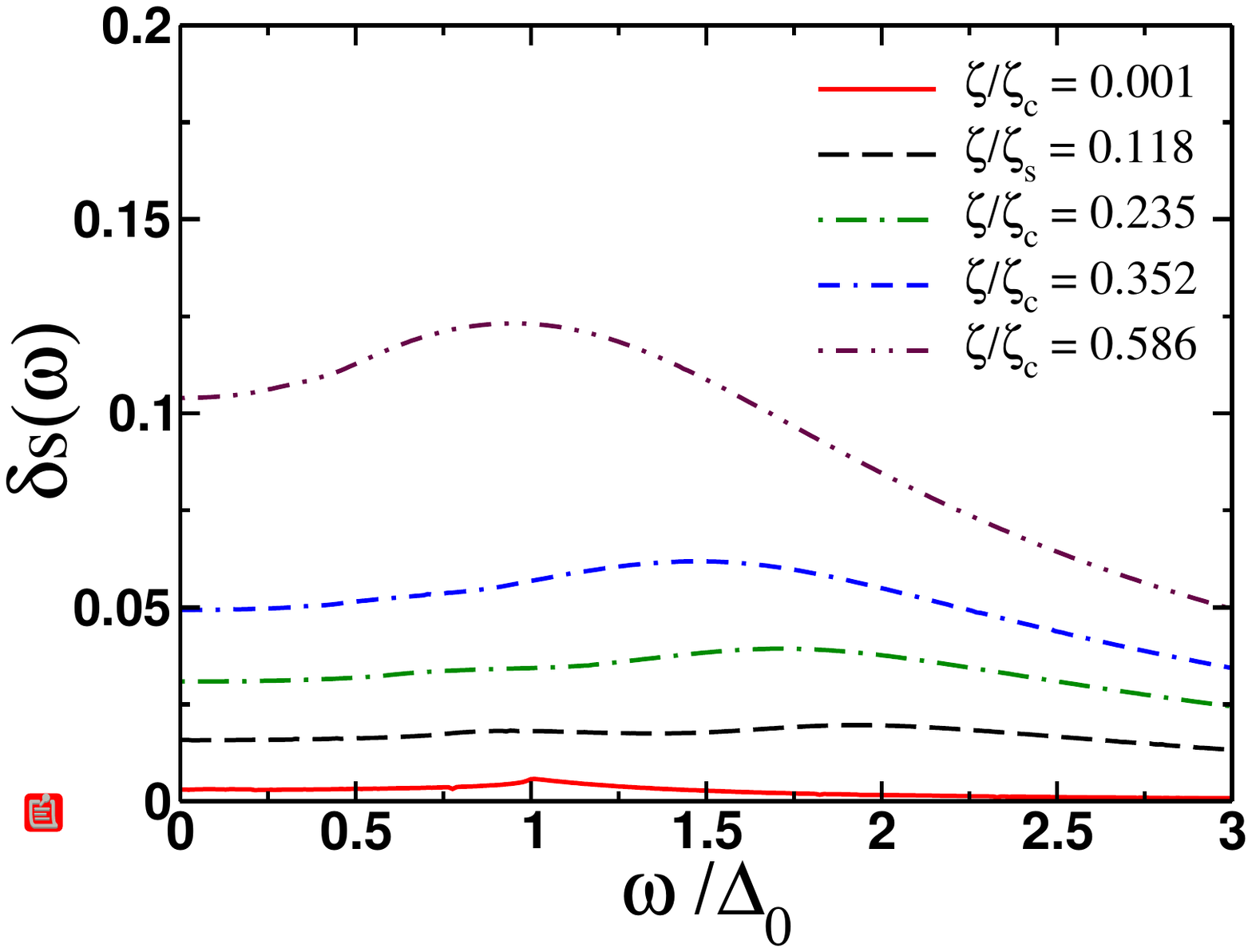}
\caption{Frequency dependence of the function $\delta s(\omega)$, Eq. (\ref{sbards}), for various values of disorder scattering rate $\zeta=1/\tau_{\textrm{s}}\Delta_0$. This function plays an important role as it accounts for the compensation of the electronic polarizability term $2\nu_F\Phi_\omega$ by the correction due to the charge re-distribution in the Poisson equation (\ref{Poisson}). In the limit $\zeta\to 0$ this function vanishes identically signalling the perfect compensation in a superconductor without spin-flip scattering. However, for finite values of $\tau_{\textrm{s}}^{-1}$ this function is nonzero which means that there will be a finite contribution to the dispersion of the superconducting plasmon from the electronic polarization effects.}
\label{Fig-sw}
\end{figure}

Let us now consider the case when $\omega\not=0$ while keeping terms linear in $Dq^2$. This situation describes the collective mode, which is equivalent to a plasmon mode in the normal state.  By construction, this mode is associated with charge re-distribution and, therefore, it is gapped since there must be an energy cost associated with it due to the Coulomb interactions.  The minimal frequency for the superconducting plasmon excitation will then be described by the solution of the following non-linear equation
\beg\label{Plasmon}
\begin{split}
\left[\begin{matrix} a(\omega) & -i\tilde{a}(\omega) \\
i\tilde{a}(\omega) & s(\omega)\end{matrix}\right]\left(
\begin{matrix}
\delta\Delta_\omega^T \\ 
\Phi_\omega
\end{matrix}
\right)=Dq^2\left[
\begin{matrix} b(\omega) & \tilde{b}(\omega) \\ -\tilde{d}(\omega) & d(\omega)\end{matrix}
\right]\left(
\begin{matrix}
\delta\Delta_\omega^T \\ 
\Phi_\omega
\end{matrix}
\right).
\end{split}
\en
\begin{figure}
\includegraphics[width=0.85\linewidth]{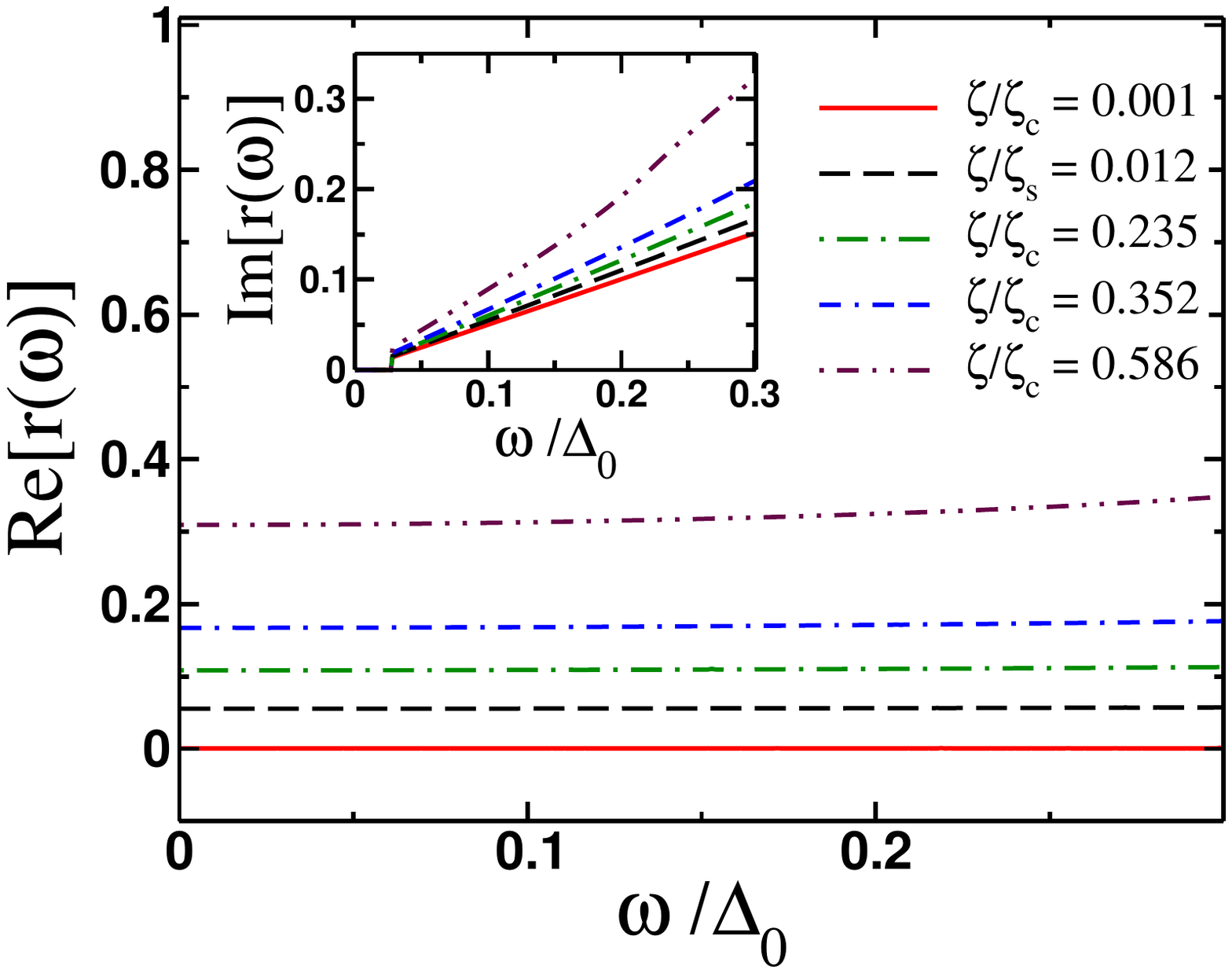}
\caption{Frequency dependence of the real and imaginary (inset) of the function $r(\omega)=a(\omega)s(\omega)-\tilde{a}^2(\omega)$, Eq. (\ref{Plasmon}), for various values of disorder scattering rate $\zeta=1/\tau_{\textrm{s}}\Delta_0$. Notably, the real part of the function $r(\omega)$ remains essentially constant, while the imaginary part of $r(\omega)$ shows approximately linear dependence on frequency provided it exceeds certain threshold value.}
\label{Fig-Detw}
\end{figure}
Two comments are in order. It proves convenient to represent $s(\omega)=\overline{s}(\omega)+\delta s(\omega)$ (see Eq. (\ref{sbards}) in Appendix A), where function $\delta s(\omega)$ reduces to the integral over $\eps$ of the first term on the right hand side of (\ref{dGKPoten}) in the limit $\tau_{\textrm{s}}\to\infty$. Then one finds that function $\delta s(\omega)$ vanishes identically in the limit $\tau_{\textrm{s}}\to\infty$. However, for nonzero scattering rate $\tau_{\textrm{s}}$ this function remains nonzero albeit small and it also displays a weak dependence on frequency as shown in Fig. \ref{Fig-sw}. This means that in the Poisson equation there remains a finite contribution from the electronic polarizability due to the pair breaking processes. Secondly, in the limit $\tau_{\textrm{s}}\to\infty$ and for $\omega\ll \Delta_0$ the determinant of the matrix on the left hand side of equation (\ref{Plasmon}) also vanishes identically (see Appendix A). Thus, here we encounter a similar situation as before: the superconducting plasmon mode can only be excited at finite values of momentum which is also given by $\tilde{q}_{\textrm{th}}\sim 1/(\tau_{\textrm{s}}\Delta_0)$. We proceed to note that equation (\ref{Plasmon}) can be further simplified by taking into that \emph{(i)} according to our numerical analysis of the functions entering into (\ref{Poisson}) it obtains $\tilde{b}(\omega)\approx \tilde{d}(\omega)$ and also $a(\omega)d(\omega)+b(\omega)s(\omega)\gg -2i\tilde{a}(\omega)\tilde{b}(\omega)$;
\emph{(ii)} $d(\omega)\approx-1/{\omega}_{\textrm{0}}$ where ${\omega}_{\textrm{0}}=8\pi^2e^2\nu_FD$ and \emph{(iii)} for the case of small scattering rate $1/\tau_{\textrm{s}}$ we also find function $r(\omega)=a(\omega)s(\omega)-\tilde{a}^2(\omega)$ to be weakly dependent of $\omega$ provided that $\omega/\Delta_0\ll 1$, Fig. \ref{Fig-Detw}.  As a result equation which determines the frequency of the superconducting plasmon mode reads
\beg\label{Eq4Plasmon}
r(\omega)=\frac{Dq^2}{\omega_0}\left[\omega_0b(\omega)s(\omega)-a(\omega)\right].
\en
From this equation, it is straightforward to find that in the limit $\tau_{\textrm{s}}\to\infty$ we have $a(\omega)\propto \omega^2$, $b(\omega)\approx1/4\Delta_0$ and $s(\omega)\propto4\Delta_0^2$. Furthermore, in this limit $r(\omega)\to 0$ and one obtains for the frequency of the superconducting plasmon $\omega_{\textrm{p}}\approx\sqrt{\omega_0\Delta_0}$ \cite{Kulik1981,KamenevBook}.
However, in the general case equation (\ref{Eq4Plasmon}) does not have a solution for $q=0$ for as we have already noted function $r(\omega)\not=0$ when $\tau_{\textrm{s}}^{-1}\not=0$, Fig. \ref{Fig-Detw}. Based on the results shown in Fig. \ref{Fig-Detw},  we can approximate $r(\omega)\approx1/\tau_{\textrm{s}}\Delta$ and also $b(\omega\ll\Delta_0)\approx b(0)$. It then follows
\beg\label{wp}
\omega_{\textrm{p}}(q)\simeq\sqrt{\omega_0\Delta}\left(1-\frac{q_0^2}{q^2}\right)^{1/2}\vartheta(q-q_0),
\en
where $q_0\sim\sqrt{1/D\tau_{\textrm{s}}}$. Thus, we found that the superconducting plasmon mode becomes gapless in the presence of the pair breaking scattering. The damping of this mode can be estimated using the limit $\tau_{\textrm{s}}\to\infty$, 
which yields $\gamma_{\textrm{p}}\sim Dq^2\log(\Delta/Dq^2)$.

\begin{figure}
\includegraphics[width=0.85\linewidth]{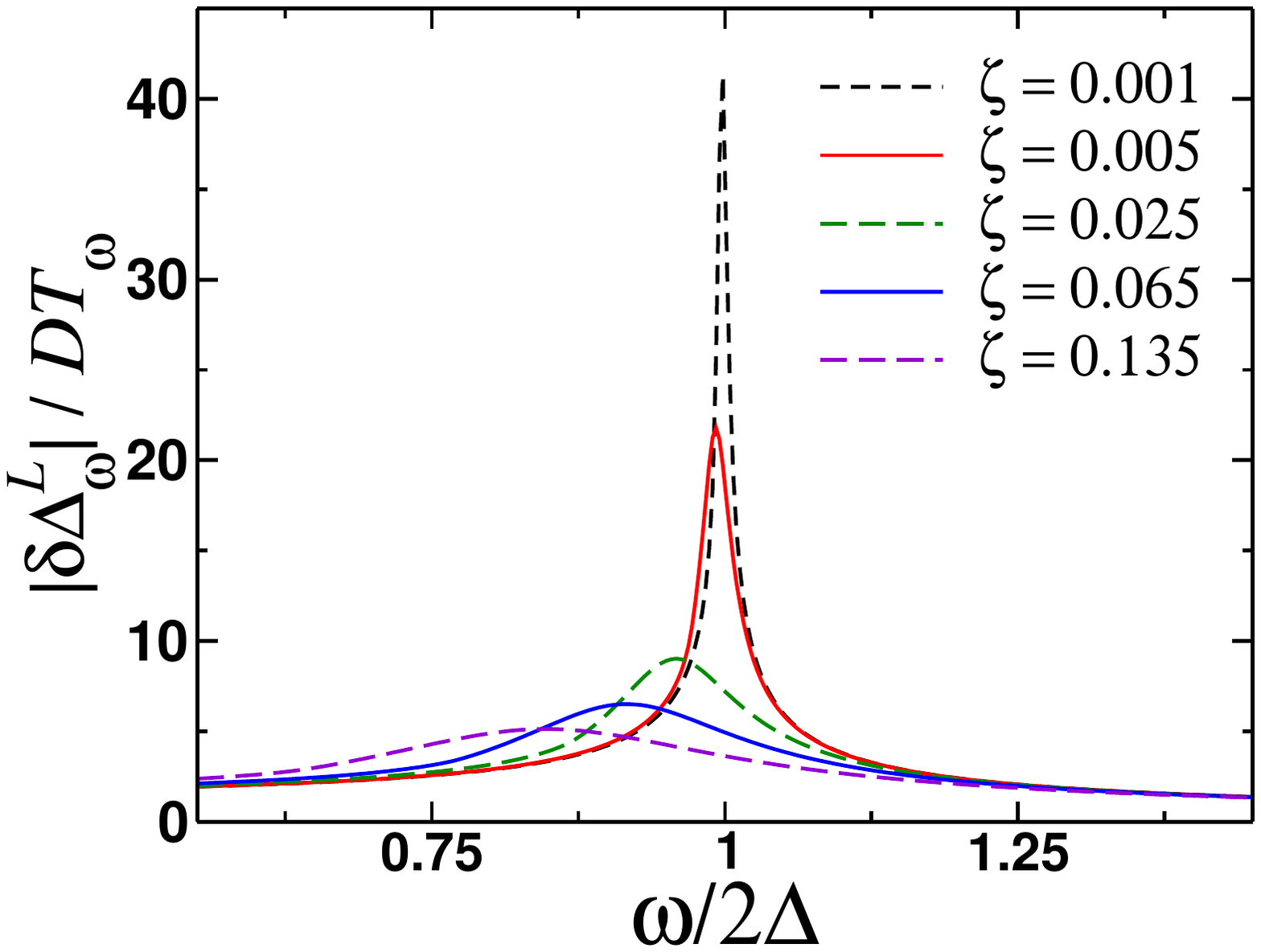}
\caption{Amplitude of the Higgs mode $|\delta\Delta_\omega^L|$ (in the units of $D{\cal T}_\omega$). 
Here $\omega=\Omega_{\nu+\mu}$, $\zeta=1/(\tau_{\textrm{s}}\Delta_0)$, $\Delta_0$ is the pairing gap in a clean superconductor and $\Delta$ is a pairing gap in the presence of paramagnetic impurities. The value of the coupling constant $\lambda=0.105$ and  
$\Delta_0/W=0.016$, where $W$ is the band width. The critical value of $\zeta$ is given by $2/(\tau_{\textrm{s}}\Delta_0)=1$ or $\zeta_{\textrm{c}}=0.855$ for our choice for the values of the coupling constant and ultraviolet cutoff. We observe that with an increase in magnetic scattering the amplitude of the resonant Higgs mode along with the resonant frequency are suppressed. While the broadening of the Higgs mode resonance with an increase in scattering is expected,  contrary to our intuition, the value of resonant frequency shifts below $2\Delta$. }
\label{Fig-NewNew}
\end{figure}

\section{Amplitude Higgs mode}
We add the discussion of the amplitude Higgs mode (also known in the literature as Schmid mode \cite{Kulik1981}) here mostly for completeness, since the main results which we will present are conceptually very similar to the ones reported in our previous publication \cite{Yantao2023}. Specifically, in Ref. \cite{Yantao2023} we have computed the value of the resonant amplitude Higgs mode under assumption that paramagnetic impurities were weak. On a technical level that approximation was equivalent to neglecting the second term ($\propto \tau_{\textrm{s}}^{-1}$) on the right hand side of the linearized Usadel equation (\ref{LinUsadel}). We will lift this assumption here and analyze the amplitude Higgs mode including the terms which were previously neglected. Since this calculation will be very similar to the one described in Ref. \cite{Yantao2023}, here we are going to provide the key expressions which one may need in order to follow (or reproduce) the calculation.
Clearly, in the expression for $\delta\hat{\Delta}$ we need to keep the longitudinal part only. Consequently, we write
\beg\label{dG}
\delta\hat{G}_{\eps\eps'}^{(\alpha)}=\delta G_{\eps\eps'}^{(\alpha)}\hat{\Xi}_3+\delta F_{\eps\eps'}^{(\alpha)}\hat{\Xi}_2, \quad \delta\hat{\Delta}(\omega)=\delta\Delta_\omega^L\hat{\Xi}_2.
\en
The value of the correction to the order parameter is given by 
\beg\label{Delta1}
\delta{\Delta}_\omega^L=
\frac{\pi\lambda}{2}\int\limits_{-\infty}^{\infty}\frac{d\eps}{2\pi}\textrm{Tr}\left\{-\hat{\Xi}_2\delta\hat{G}_{\eps,\eps-\omega}^K\right\}.
\en
We will simplify our subsequent analysis of the linearized Usadel equation by ignoring \emph{(i)} the gradient term and \emph{(ii)} the scalar potential. 
The first approximation is justified by the fact that the amplitude mode is gapped. The second approximation is justified by our choice for the ground state configuration of the $Q$-matrix, which guarantees that the Poisson equation and self-consistency equation for $\delta\Delta_\omega^L$ will be decoupled from each other.
The Keldysh function can still be represented as (\ref{dGKAB}). The regular contribution $\delta G_{\textrm{reg}}^K(\eps,\eps')$ is expressed in the terms of the retarded and advanced components which evaluate to
\beg\label{dGRAHiggs}
\begin{split}
\delta G_{\eps\eps'}^{\alpha}&=\frac{\left(Z_{\eps\eps'}^{\alpha}+\gamma[A_{\eps\eps'}^{(+)}]^{\alpha}\right)\rho_3^{\alpha}(\eps,\eps')-\gamma B_{\eps\eps'}^{\alpha}\rho_2^{\alpha}(\eps,\eps')}{\left(Z_{\eps\eps'}^{\alpha}-\gamma[A_{\eps\eps'}^{(-)}]^{\alpha}\right)\left(Z_{\eps\eps'}^{\alpha}+\gamma[A_{\eps\eps'}^{(+)}]^{\alpha}\right)-(\gamma B_{\eps\eps'}^{\alpha})^2}, \\ 
\delta F_{\eps\eps'}^{\alpha}&=\frac{\left(Z_{\eps\eps'}^\alpha-\gamma[A_{\eps\eps'}^{(-)}]^\alpha\right)\rho_2^\alpha(\eps,\eps')-\gamma B_{\eps\eps'}^\alpha\rho_3^\alpha(\eps,\eps')}{\left(Z_{\eps\eps'}^\alpha-\gamma[A_{\eps\eps'}^{(-)}]^\alpha\right)\left(Z_{\eps\eps'}^\alpha+\gamma[A_{\eps\eps'}^{(+)}]^\alpha\right)-(\gamma B_{\eps\eps'}^\alpha)^2}
\end{split}
\en
with $\gamma=i/2\tau_{\textrm{s}}$, $\alpha=R,A$ and the expressions for all the functions which appear here are listed in Appendix \ref{AppendixA}. The corresponding expressions listed in Ref. \cite{Yantao2023} can be recovered from (\ref{dGRAHiggs}) by setting $\gamma=0$.

For the components of the anomalous part of the Keldysh function we found
\beg\label{dGanK}
\begin{split}
\delta G_{\textrm{an}}^K(\eps,\eps')&=\frac{\left({Z}_{\eps\eps'}+\gamma{A}_{\eps\eps'}^{(+)}\right){\cal P}_3(\eps,\eps')-\gamma{B}_{\eps\eps'}{\cal P}_2(\eps,\eps')}{\left({Z}_{\eps\eps'}-\gamma{A}_{\eps\eps'}^{(-)}\right)\left({Z}_{\eps\eps'}+\gamma{A}_{\eps\eps'}^{(+)}\right)-(\gamma{B}_{\eps\eps'})^2}, \\ 
\delta F_{\textrm{an}}^K(\eps,\eps')&=\frac{\left({Z}_{\eps\eps'}-\gamma{A}_{\eps\eps'}^{(-)}\right){\cal P}_2(\eps,\eps')-\gamma{B}_{\eps\eps'}{\cal P}_3(\eps,\eps')}{\left({Z}_{\eps\eps'}-\gamma{A}_{\eps\eps'}^{(-)}\right)\left({Z}_{\eps\eps'}+\zeta{A}_{\eps\eps'}^{(+)}\right)-(\gamma{B}_{\eps\eps'})^2}.
\end{split}
\en
Here functions ${\cal P}_{2,3}(\eps,\eps')$ are obtained from the matrix elements of 
\beg\label{BigP}
\check{\cal P}_{\eps\eps'}=2\pi\sum\limits_{\mu\nu}\left[\check{R}_{\cal A}(\eps,\eps')+\check{R}_{\Delta}(\eps,\eps')\right]\delta(\eps-\eps'-\Omega_{\nu+\mu}).
\en
(see Appendix A for details). Inserting these expressions into the self-consistency equation (\ref{Delta1}) and separating the terms which are proportional to $\delta\Delta_\omega^L$ yields the following equation:
\beg\label{Eq4Delta1}
\left[C_{\textrm{reg}}(\omega)+C_{\textrm{an}}(\omega)\right]\delta\Delta_\omega^L=2iD{\cal T}_\omega\left[B_{\textrm{reg}}(\omega)+B_{\textrm{an}}(\omega)\right].
\en
The results of the numerical calculation of the dependence $|\delta\Delta_\omega^L|$ on $\omega$ from equation (\ref{Eq4Delta1}) are presented in Fig. \ref{Fig-NewNew}. Expectantly, we find that the resonant frequency of the amplitude Higgs mode shifts to smaller values with an increase in the spin-flip scattering. It is worth noting \emph{(i)} quite substantial decrease in the values of $|\delta\Delta_\omega^L|$ itself as the value of $1/\tau_{\textrm{s}}$ increases and \emph{(ii)} broadening of the resonance, which indicates that the experimental observation of this resonance may be problematic when the concentration of paramagnetic impurities is not necessarily small, $\zeta\geq 0.2\zeta_c$. Both of these results provide an important correction to our earlier perturbative ones \cite{Yantao2023}. Lastly, we would also like to mention that by retaining terms $Dq^2$ in the expressions above, we found that the amplitude Higgs mode is diffusive, $\omega_{\textrm{amp}}\sim\omega_{\textrm{Higgs}}+Dq^2$.

\section{Conclusions}
In this paper we have revisited an old problem of collective modes in disordered superconductors. We have considered a new aspect of that problem: how the presence of the the spin-flip scattering due to paramagnetic impurities affects the excitation spectrum of the collective modes. In agreement with the earlier studies we found that under the influence of the external terahertz field paramagnetic impurities produce a red shift in the resonant frequency of the amplitude Higgs mode \cite{Dzero2023-Disorder,Yantao2023}. Furthermore, we found that excitation of the phase and superconducting plasom modes require a minimum momentum threshold which is determined by the mean-free path between the two consecutive spin-flip scattering processes. As a result of this, superconducting plasmon mode becomes gapless which is in contrast with the case of the plasmon mode in diffusive superconductors where this mode remains gapped. 

Finally, we would like to comment on the possibility of the experimental verification of our results. We think that the red shift in the resonant frequency of the amplitude mode should be directly observable using the experimental setup similar to the one in Ref. \cite{Armitage2023}. Another possibility would be to probe this red shift in the inverse Faraday effect by measuring the frequency dependence of the induced $dc$-magnetization, which is predicted to have a minimum at the frequency corresponding to the resonance of the amplitude Higgs mode \cite{dzero2024ife}. As for the spectrum of the Carlson-Goldman and superconducting plasmon modes it is not immediately clear to us at present how these modes can be measured using monochromatic pulses. However, if superconductor is subjected to a circular polarized light these modes should also contribute to either ac- or dc-response functions provided that the wave vector for the external field is of the order of $1/\sqrt{D\tau_{\textrm{s}}}$. With this being said, we think that these modes can also be probed by measuring the $I$-$V$ characteristics of the Josephson junction by applying a magnetic field along the plane of the junction \cite{Carlson1975,Kulik1981} or using the near-field setup. In particular, we expect that there should be a nonzero value of the magnetic field required to excite these modes.

\section{Acknowledgements} Useful discussion with Alex Levchenko is gratefully acknowledged. This work was financially  supported by the National Science Foundation grants NSF-DMR-2002795 (M.D.) and NSF-DMR-1904315 (Y.L.)

\begin{appendix}
\begin{widetext}
\section{Auxiliary expressions}\label{AppendixA}
In this Section we define several auxiliary functions which are used in the main text. 
\paragraph{Anderson-Bogoliubov mode.} The functions which enter into expressions (\ref{dGFanFin}) are defined according to
\beg\label{ABs}
\begin{split}
&{\cal A}_{\eps\eps'}^{R(A)}=1+g_{\eps}^{R(A)}g_{\eps'}^{R(A)}-f_{\eps}^{R(A)}f_{\eps'}^{R(A)}, \quad 
\overline{\cal A}_{\eps\eps'}^{R(A)}=1-g_{\eps}^{R(A)}g_{\eps'}^{R(A)}+f_{\eps}^{R(A)}f_{\eps'}^{R(A)}, \\
&{\cal A}_{\eps\eps'}^{K}=1+g_{\eps}^{R}g_{\eps'}^{A}-f_{\eps}^{R}f_{\eps'}^{A}, \quad 
\overline{\cal A}_{\eps\eps'}^{K}=1-g_{\eps}^{R}g_{\eps'}^{A}+f_{\eps}^{R}f_{\eps'}^{A}, \\
&{\cal B}_{\eps\eps'}^{R(A)}=g_{\eps}^{R(A)}f_{\eps'}^{R(A)}-f_{\eps}^{R(A)}g_{\eps'}^{R(A)}, \quad {\cal B}_{\eps\eps'}^{K}=g_{\eps}^{R}f_{\eps'}^{A}-f_{\eps}^{R}g_{\eps'}^{A}.
\end{split}
\en
We note that function ${\cal A}_{\eps\eps'}^{R(A)}$ becomes independent on $\eps$ when $\eps'=\eps$, while $\overline{\cal A}_{\eps\eps}^{R(A)}$ as well as ${\cal B}_{\eps\eps}^{R(A)}$ become equal to zero.
\paragraph{{Carlson}-Goldman mode.} In addition to the functions we introduced above, we also consider
\beg\label{Not4dGKFin}
\begin{split}
{n}_{\eps\eps'}&=\frac{(t_\eps-t_{\eps'}){\cal B}_{\eps\eps'}^K}{(\eta_\eps^R+\eta_{\eps'}^A)^2}+\frac{{\cal B}_{\eps\eps'}^Rt_{\eps'}}{(\eta_\eps^R+\eta_{\eps'}^R)^2}-\frac{t_\eps{\cal B}_{\eps\eps'}^A}{(\eta_\eps^A+\eta_{\eps'}^A)^2}, \quad \tilde{n}_{\eps\eps'}=\frac{(t_\eps-t_{\eps'}){\cal B}_{\eps\eps'}^K}{\eta_\eps^R+\eta_{\eps'}^A}+\frac{{\cal B}_{\eps\eps'}^Rt_{\eps'}}{\eta_\eps^R+\eta_{\eps'}^R}-\frac{t_\eps{\cal B}_{\eps\eps'}^A}{\eta_\eps^A+\eta_{\eps'}^A}, \\ 
\tilde{m}_{\eps\eps'}&=\frac{t_\eps\left({\cal A}_{\eps\eps}^K{\cal B}_{\eps\eps'}^A+{\cal B}_{\eps\eps}^K{\cal A}_{\eps\eps'}^A\right)}{(\eta_\eps^R+\eta_{\eps'}^A)(\eta_\eps^A+\eta_{\eps'}^A)}, \quad 
m_{\eps\eps'}=\frac{t_\eps\left({\cal A}_{\eps\eps}^K{\cal B}_{\eps\eps'}^A-{\cal B}_{\eps\eps}^K{\cal A}_{\eps\eps'}^A\right)}{(\eta_\eps^R+\eta_{\eps'}^A)(\eta_\eps^A+\eta_{\eps'}^A)}, \\
{p}_{\eps\eps'}&=\frac{(t_\eps-t_{\eps'}){\cal A}_{\eps\eps'}^K}{(\eta_\eps^R+\eta_{\eps'}^A)^2}+\frac{{\cal A}_{\eps\eps'}^Rt_{\eps'}}{(\eta_\eps^R+\eta_{\eps'}^R)^2}-\frac{t_\eps{\cal A}_{\eps\eps'}^A}{(\eta_\eps^A+\eta_{\eps'}^A)^2}, \quad \tilde{p}_{\eps\eps'}=\frac{(t_\eps-t_{\eps'}){\cal A}_{\eps\eps'}^K}{\eta_\eps^R+\eta_{\eps'}^A}+\frac{{\cal A}_{\eps\eps'}^Rt_{\eps'}}{\eta_\eps^R+\eta_{\eps'}^R}-\frac{t_\eps{\cal A}_{\eps\eps'}^A}{\eta_\eps^A+\eta_{\eps'}^A}, \\
{s}_{\eps\eps'}&=\frac{t_\eps\left({\cal A}_{\eps\eps}^K{\cal A}_{\eps\eps'}^A+{\cal B}_{\eps\eps}^K{\cal B}_{\eps\eps'}^A\right)}{(\eta_\eps^R+\eta_{\eps'}^A)(\eta_\eps^A+\eta_{\eps'}^A)}, \quad \tilde{s}_{\eps\eps'}=\frac{t_\eps\left({\cal A}_{\eps\eps}^K{\cal A}_{\eps\eps'}^A-{\cal B}_{\eps\eps}^K{\cal B}_{\eps\eps'}^A\right)}{(\eta_\eps^R+\eta_{\eps'}^A)(\eta_\eps^A+\eta_{\eps'}^A)}.
\end{split}
\en
All these functions must be computed for $\eps'=\eps-\omega$. When $\omega=0$ we then observe that all these functions are zero.

Coefficients in the system of linear equations (\ref{SelfCC}) are defined by 
\beg\label{tabw}\nonumber
\begin{split}
&\tilde{a}(\omega)=\frac{1}{4}\int\limits_{-\infty}^\infty\left[\frac{(t_\eps-t_{\eps-\omega}){\cal B}_{\eps,\eps-\omega}^K}{\eta_\eps^R+\eta_{\eps-\omega}^A}+\frac{{\cal B}_{\eps,\eps-\omega}^Rt_{\eps-\omega}}{\eta_\eps^R+\eta_{\eps-\omega}^R}-\frac{t_\eps{\cal B}_{\eps,\eps-\omega}^A}{\eta_\eps^A+\eta_{\eps-\omega}^A}\right]d\eps, \\
&\tilde{d}(\omega)=\frac{1}{4}\int\limits_{-\infty}^\infty\left[\frac{(t_\eps-t_{\eps-\omega}){\cal B}_{\eps,\eps-\omega}^K}{(\eta_\eps^R+\eta_{\eps-\omega}^A)^2}+\frac{{\cal B}_{\eps,\eps-\omega}^Rt_{\eps-\omega}}{(\eta_\eps^R+\eta_{\eps-\omega}^R)^2}-\frac{t_\eps{\cal B}_{\eps,\eps-\omega}^A}{(\eta_\eps^A+\eta_{\eps-\omega}^A)^2}+\frac{t_\eps\left({\cal A}_{\eps\eps}^K{\cal B}_{\eps,\eps-\omega}^A+{\cal B}_{\eps\eps}^K\overline{\cal A}_{\eps,\eps-\omega}^A\right)}{(\eta_\eps^R+\eta_{\eps-\omega}^A)(\eta_\eps^A+\eta_{\eps-\omega}^A)}\right]d\eps, \\
&\tilde{b}(\omega)=\frac{1}{4}\int\limits_{-\infty}^\infty\left[\frac{(t_\eps-t_{\eps-\omega}){\cal B}_{\eps,\eps-\omega}^K}{(\eta_\eps^R+\eta_{\eps-\omega}^A)^2}+\frac{{\cal B}_{\eps,\eps-\omega}^Rt_{\eps-\omega}}{(\eta_\eps^R+\eta_{\eps-\omega}^R)^2}-\frac{t_\eps{\cal B}_{\eps,\eps-\omega}^A}{(\eta_\eps^A+\eta_{\eps-\omega}^A)^2}+\frac{t_\eps\left({\cal A}_{\eps\eps}^K{\cal B}_{\eps,\eps-\omega}^A-{\cal B}_{\eps\eps}^K{\cal A}_{\eps,\eps-\omega}^A\right)}{(\eta_\eps^R+\eta_{\eps-\omega}^A)(\eta_\eps^A+\eta_{\eps-\omega}^A)}\right]d\eps, \\
&s(\omega)=1-\frac{1}{4}\int\limits_{-\infty}^\infty\left[\frac{(t_\eps-t_{\eps-\omega})\overline{\cal A}_{\eps,\eps-\omega}^K}{\eta_\eps^R+\eta_{\eps-\omega}^A}+\frac{\overline{\cal A}_{\eps,\eps-\omega}^Rt_{\eps-\omega}}{\eta_\eps^R+\eta_{\eps-\omega}^R}-\frac{t_\eps\overline{\cal A}_{\eps,\eps-\omega}^A}{\eta_\eps^A+\eta_{\eps-\omega}^A}\right]d\eps, \\
&{d}(\omega)=-\frac{1}{8\pi e^2\nu_FD}+\frac{i}{4}\int\limits_{-\infty}^\infty d\eps\left[\frac{(t_\eps-t_{\eps-\omega})\overline{\cal A}_{\eps,\eps-\omega}^K}{(\eta_\eps^R+\eta_{\eps-\omega}^A)^2}+
\frac{\overline{\cal A}_{\eps,\eps-\omega}^Rt_{\eps-\omega}}{(\eta_\eps^R+\eta_{\eps-\omega}^R)^2}-
\frac{t_\eps\overline{\cal A}_{\eps,\eps-\omega}^A}{(\eta_\eps^A+\eta_{\eps-\omega}^A)^2}+\frac{t_\eps\left({\cal A}_{\eps\eps}^K\overline{\cal A}_{\eps,\eps-\omega}^A+{\cal B}_{\eps\eps}^K{\cal B}_{\eps,\eps-\omega}^A\right)}{(\eta_\eps^R+\eta_{\eps-\omega}^A)(\eta_\eps^A+\eta_{\eps-\omega}^A)}\right].
\end{split}
\en
The last three functions vanish identically in the limit $\omega\to 0$. This means that the corresponding integrals are convergent at $|\eps|\gg \omega$. 

For our subsequent discussion in main text, it will be useful to represent $s(\omega)$ as a sum of two contributions, $s(\omega)=\overline{s}(\omega)+\delta s(\omega)$, where
\beg\label{sbards}
\begin{split}
\overline{s}(\omega)&=-\frac{1}{2}\int\limits_{-\infty}^\infty\left[
\frac{(t_\eps-t_{\eps-\omega})f_\eps^Rf_{\eps-\omega}^A}{\eta_\eps^R+\eta_{\eps-\omega}^A}+\frac{f_\eps^Rf_{\eps-\omega}^Rt_{\eps-\omega}}{\eta_\eps^R+\eta_{\eps-\omega}^R}-\frac{t_\eps f_\eps^Af_{\eps-\omega}^A}{\eta_\eps^A+\eta_{\eps-\omega}^A}
\right]d\eps, \\
\delta s(\omega)&=1-\frac{1}{4}\int\limits_{-\infty}^\infty\left(\frac{(t_\eps-t_{\eps-\omega})(1-g_\eps^Rg_{\eps-\omega}^A-f_\eps^Rf_{\eps-\omega}^A)}{\eta_\eps^R+\eta_{\eps-\omega}^A}+\frac{(1-g_\eps^Rg_{\eps-\omega}^R-f_\eps^Rf_{\eps-\omega}^R)t_{\eps-\omega}}{\eta_\eps^R+\eta_{\eps-\omega}^R}-\frac{t_\eps(1-g_{\eps}^Ag_{\eps-\omega}^A-f_{\eps}^Af_{\eps-\omega}^A)}{\eta_\eps^A+\eta_{\eps-\omega}^A}\right]d\eps.
\end{split}
\en
The reason behind such a re-arrangement of terms lies in the fact that in the limit when $\tau_{\textrm{s}}\to\infty$, it is easy to prove (see the main text below Eq. (\ref{dGKPoten})) $\delta s(\omega)\to 0$ independent on the value of $\omega$. We have computed the frequency dependence of this function numerically and present our results in Fig. \ref{Fig-sw} in the main text.
Finally, we would also like to separately consider function $r(\omega)=a(\omega)s(\omega)-\tilde{a}^2(\omega)$ which appears on the left hand side of the equation (\ref{Plasmon}). 

\paragraph{Amplitude Higgs mode.} Functions which appear in expressions (\ref{dGRAHiggs}) are defined as follows:
\beg\label{AuxHiggs}\nonumber
\begin{split}
[A_{\eps\eps'}^{(\pm)}]^{R(A)}&=1\pm\left(g_{\eps}^{R(A)}g_{\eps'}^{R(A)}+f_{\eps}^{R(A)}f_{\eps'}^{R(A)}\right), \quad 
B_{\eps\eps'}^{R(A)}=g_{\eps}^{R(A)}f_{\eps'}^{R(A)}+g_{\eps}^{R(A)}f_{\eps'}^{R(A)},\quad Z_{\eps\eps'}^{R(A)}=\eta_\eps^{R(A)}+\eta_{\eps'}^{R(A)}, \\
\check{\rho}_{\cal A}(\eps,\eps')&+\check{\rho}_{\Delta}(\eps,\eps')=\rho_3(\eps,\eps')\check{\Xi}_3+\rho_2(\eps,\eps')\check{\Xi}_2, \quad 
\check{\rho}_{\Delta}=\delta\check{\Delta}(\eps-\eps')-\check{\Lambda}_\eps\delta\check{\Delta}(\eps-\eps')\check{\Lambda}_{\eps'}, \\
\check{\rho}_{\cal A}&=\check{\Xi}_3\left(\hat{\Lambda}_{\eps}+\check{\Lambda}_{\eps'}\right)\check{\Xi}_3-
\check{\Lambda}_{\eps}\check{\Xi}_3\left(\check{\Lambda}_{\eps}+\check{\Lambda}_{\eps'}\right)\check{\Xi}_3\check{\Lambda}_{\eps'}
\end{split}
\en
The matrices entering into the last three expressions are defined in both Nambu and Keldysh spaces. We continue with the definitions of the functions which enter into equations (\ref{dGanK}) are given by
\beg\label{P3P2}\nonumber
\begin{split}
&\hat{\Lambda}_\eps^R\left(\hat{\cal P}_{\eps\eps'}^K+t_\eps\hat{\cal P}_{\eps\eps'}^A-\hat{\cal P}_{\eps\eps'}^Rt_{\eps'}\right)={\cal P}_3(\eps,\eps')\hat{\Xi}_3+{\cal P}_2(\eps,\eps')\hat{\Xi}_2, \\
&{A}_{\eps\eps'}^{(\pm)}=1\pm\left(g_{\eps}^Rg_{\eps'}^A+f_{\eps}^Rf_{\eps'}^A\right), ~{B}_{\eps\eps'}=g_{\eps}^Rf_{\eps'}^A+g_{\eps}^Rf_{\eps'}^A, \quad {Z}_{\eps\eps'}=\eta_\eps^R+\eta_{\eps'}^A.
\end{split}
\en
Finally, we define functions which appear in Eq. (\ref{Eq4Delta1}). Function $C_{\textrm{reg}}(\omega)$ is defined as $C_{\textrm{reg,an}}(\omega)=\int_{-\infty}^\infty d\eps C_{\textrm{reg,an}}(\eps,\eps-\omega)$ with 
\beg\label{Creg}\nonumber
\begin{split}
C_{\textrm{reg}}(\eps,\eps')&=\frac{\left[\left(Z_{\eps\eps'}^R-\gamma[A_{\eps\eps'}^{(-)}]^R\right)[A_{\eps\eps'}^{(+)}]^R-\gamma[B_{\eps\eps'}^R]^2\right]t_{\eps'}}{\left(Z_{\eps\eps'}^R-\gamma[A_{\eps\eps'}^{(-)}]^R\right)\left(Z_{\eps\eps'}^R+\gamma[A_{\eps\eps'}^{(+)}]^R\right)-[\gamma B_{\eps\eps'}^R]^2}-\frac{t_\eps\left[\left(Z_{\eps\eps'}^A-\gamma[A_{\eps\eps'}^{(-)}]^A\right)[A_{\eps\eps'}^{(+)}]^A-\gamma[B_{\eps\eps'}^A]^2\right]}{\left(Z_{\eps\eps'}^A-\eta_s[A_{\eps\eps'}^{(-)}]^A\right)\left(Z_{\eps\eps'}^A+\gamma[A_{\eps\eps'}^{(+)}]^A\right)-[\gamma B_{\eps\eps'}^A]^2}-\left(\frac{f_\eps^R-f_\eps^A}{\Delta}\right)t_\eps.
\end{split}
\en
The last term appears after we replace the coupling constant with the self-consistency condition in equilibrium.  As we have already mentioned at several instances, in the limit $\gamma\to 0$ we readily recover the corresponding expressions in Ref. \cite{Yantao2023}. Function $C_{\textrm{an}}(\Omega)$ is defined in complete analogy with $C_{\textrm{reg}}(\eps,\eps')$: 
\beg\label{Canom}\nonumber
\begin{split}
C_{\textrm{an}}(\eps,\eps')&=\frac{\left[\left(Z_{\eps\eps'}-\gamma[A_{\eps\eps'}^{(-)}]\right)A_{\eps\eps'}^{(+)}-\gamma B_{\eps\eps'}^2\right](t_\eps-t_{\eps'})}{\left(Z_{\eps\eps'}-\gamma A_{\eps\eps'}^{(-)}\right)\left(Z_{\eps\eps'}+\gamma A_{\eps\eps'}^{(+)}\right)-[\gamma B_{\eps\eps'}]^2}.
\end{split}
\en
Functions $B_{\textrm{reg,an}}(\Omega)$ are defined as $B_{\textrm{reg,an}}(\omega)=\int_{-\infty}^\infty d\eps B_{\textrm{reg,an}}(\eps,\eps-\omega)$ with
\beg\label{Bregan}\nonumber
\begin{split}
&B_{\textrm{reg}}(\eps,\eps')=\frac{\left(Z_{\eps\eps'}^R-\gamma[A_{\eps\eps'}^{(-)}]^R\right)(g_\eps^R+g_{\eps'}^R)-\gamma(f_\eps^R+f_{\eps'}^R)B_{\eps\eps'}^R}{\left(Z_{\eps\eps'}^R-\gamma[A_{\eps\eps'}^{(-)}]^R\right)\left(Z_{\eps\eps'}^R+\gamma[A_{\eps\eps'}^{(+)}]^R\right)-[\gamma B_{\eps\eps'}^R]^2}B_{\eps\eps'}^Rt_{\eps'}-\frac{\left(Z_{\eps\eps'}^A-\gamma[A_{\eps\eps'}^{(-)}]^A\right)(g_\eps^A+g_{\eps'}^A)-\gamma(f_\eps^A+f_{\eps'}^A)B_{\eps\eps'}^A}{\left(Z_{\eps\eps'}^A-\gamma[A_{\eps\eps'}^{(-)}]^A\right)\left(Z_{\eps\eps'}^A+\gamma[A_{\eps\eps'}^{(+)}]^A\right)-[\gamma B_{\eps\eps'}^A]^2}t_\eps B_{\eps\eps'}^A, \\
&B_{\textrm{an}}(\eps,\eps')=\left\{\frac{\left(Z_{\eps\eps'}-\gamma A_{\eps\eps'}^{(-)}\right)(g_\eps^R+g_{\eps'}^A)-\gamma(f_\eps^R+f_{\eps'}^A)B_{\eps\eps'}}{\left(Z_{\eps\eps'}-\gamma A_{\eps\eps'}^{(-)}\right)\left(Z_{\eps\eps'}+\gamma A_{\eps\eps'}^{(+)}\right)-[\gamma B_{\eps\eps'}]^2}\right\}(t_\eps B_{\eps\eps'}-B_{\eps\eps'}t_{\eps'}).
\end{split}
\en
As we have already mentioned above, in the limit of small $\gamma$ we recover the corresponding expression from Ref. \cite{Yantao2023}.
\section{Expressions for the propagators in the limit $\tau_{\textrm{s}}\to\infty.$}
With the help of (\ref{etaRA}) we re-write expressions (\ref{CompsReg}) as follows
\beg\label{Simplify}
\begin{split}
&\left[\delta{G}_{\eps\eps'}^{R(A)}(q=0)\right]_{\tau_{\textrm{s}}\to\infty}=\left[\frac{{\cal B}_{\eps\eps'}^{R(A)}i\delta\Delta_\omega^T-\overline{\cal A}_{\eps\eps'}^{R(A)}\Phi_\omega}{\eta_\eps^{R(A)}+\eta_{\eps'}^{R(A)}}\right]_{\tau_{\textrm{s}}\to\infty}=
\frac{f_{\eps'}^{R(A)}-f_{\eps}^{R(A)}}{\eps+\eps'}i\delta\Delta_\omega^T+
\frac{g_{\eps}^{R(A)}+g_{\eps'}^{R(A)}}{\eps+\eps'}\Phi_\omega-\frac{2\Phi_\omega}{\eta_\eps^{R(A)}+\eta_{\eps'}^{R(A)}}, \\
&\left[\delta{F}_{\eps\eps'}^{R(A)}(q=0)\right]_{\tau_{\textrm{s}}\to\infty}=\left[\frac{{\cal A}_{\eps\eps'}^{R(A)}i\delta\Delta_\omega^T+{\cal B}_{\eps\eps'}^{R(A)}\Phi_\omega}{\eta_\eps^{R(A)}+\eta_{\eps'}^{R(A)}}\right]_{\tau_{\textrm{s}}\to\infty}=\frac{g_{\eps}^{R(A)}+g_{\eps'}^{R(A)}}{\eps+\eps'}i\delta\Delta_\omega^T+\frac{f_{\eps'}^{R(A)}-f_{\eps}^{R(A)}}{\eps+\eps'}\Phi_\omega.
\end{split}
\en
The last two terms in the expression for $\delta{G}_{\eps\eps'}^{R(A)}(q=0)$ can be further simplified (we are omitting superscipts $R(A)$ for brevity):
\beg\label{FurtherSimple}
\begin{split}
&\frac{2}{\eta_\eps+\eta_{\eps'}}-\frac{g_{\eps}+g_{\eps'}}{\eps+\eps'}=\frac{(\eta_\eps-\eta_{\eps'})(1-g_\eps g_{\eps'}+f_{\eps}f_{\eps'})}{\omega(\eps+\eps')}=\frac{g_{\eps}+g_{\eps'}}{\omega}+\frac{2\Delta(f_\eps-f_{\eps'})}{\omega(\eps+\eps')}
\end{split}
\en
and we also took into account the definition $\eps'=\eps-\omega$. Similarly, for the anomalous contribution (\ref{dGanKtaus}) we have 
\beg\label{dGanKtaus}
\begin{split}
[\delta G_{\textrm{an}}^K(\eps,\eps';q=0)]_{\tau_{\textrm{s}}\to\infty}&=(t_\eps-t_{\eps'})\left[\frac{(f_{\eps'}^A-f_{\eps}^R)}{\eps+\eps'}i\delta\Delta_\omega^T-\frac{g_\eps^R+g_{\eps'}^A}{\omega}\Phi_\omega-\frac{2\Delta(f_{\eps}^R-f_{\eps'}^A)}{\omega(\eps+\eps')}\Phi_\omega\right], \\ 
\left[\delta F_{\textrm{an}}^K(\eps,\eps';q=0)\right]_{\tau_{\textrm{s}}\to\infty}&=
(t_\eps-t_{\eps'})\left[\frac{g_\eps^R+g_{\eps'}^A}{\eps+\eps'}i\delta\Delta_\omega^T+\frac{f_{\eps'}^{A}-f_{\eps}^{R}}{\eps+\eps'}\Phi_\omega\right].
\end{split}
\en
Using the definition $\delta\hat{G}_{\eps\eps'}^K=\delta\hat{G}_{\eps\eps'}^Rt_{\eps'}-t_\eps\delta\hat{G}_{\eps\eps'}^A+\delta\hat{G}_{\textrm{an}}^K$ we recover equations (\ref{dGKPoten}) in the main text. 
\end{widetext}
\end{appendix}


\end{document}